\documentclass[a4paper,fleqn,usenatbib,useAMS]{mnras}

\usepackage{graphicx}	
\usepackage{graphics}
\usepackage{amsmath}	
\usepackage{amssymb}	
\usepackage{amsfonts}
\usepackage{multicol}        
\usepackage{bm}		
\usepackage{pdflscape}	
\usepackage{natbib}
\usepackage{epstopdf}

\usepackage[T1]{fontenc}
\usepackage{ae,aecompl}

\title[deep learning for foreground removal]{The stability of deep learning for 21cm foreground removal across various sky models and frequency-dependent systematics }

\author[T. Chen]{
  T.\,Chen$^1$\!\thanks{E-mail:~\url{tianyue.chen@epfl.ch}},
  M.\,Bianco$^1$,
  E.\,Tolley$^1$,
  M.\,Spinelli$^2$,
  D.\,Forero-Sanchez$^1$,
  J.P.\,Kneib$^1$
\\
$^1$Institute of Physics, Laboratory of Astrophysics, Ecole Polytechnique F\'{e}d\'{e}rale de Lausanne (EPFL), Observatoire de Sauverny, \\
1290 Versoix, Switzerland\\
$^2$ Institute for Particle Physics and Astrophysics, ETH Z{\"u}rich, Wolfgang Pauli Strasse 27, 8093 Z{\"u}rich, Switzerland\\
}

\begin{document}
\label{firstpage}
\pagerange{\pageref{firstpage}--\pageref{lastpage}}
\maketitle


\begin{abstract}
Deep learning (DL) has recently been proposed as a novel  approach for 21cm foreground removal. Before applying DL to real observations, it is essential to assess its consistency with established methods, its performance across various simulation models and its robustness against instrumental systematics. This study develops a commonly used U-Net  and evaluates its performance for post-reionisation foreground removal across  three distinct sky simulation models based on pure Gaussian realisations, the Lagrangian perturbation theory, and the Planck sky model. Stable outcomes across the models are achieved provided that training and testing data align with the same model. On average, the residual foreground in the U-Net reconstructed data is $\sim10\%$ of the signal across  angular scales at the considered redshift range. Comparable  results are found with traditional approaches.  However, blindly using a network trained on one model for data from another model yields inaccurate reconstructions, emphasising the need for consistent training data.   The study then introduces frequency-dependent Gaussian beams and gain drifts to the test data. The network struggles to denoise data affected by ``unexpected'' systematics without prior information.  However, after re-training consistently  with systematics-contaminated data, the network effectively restores its reconstruction accuracy. This highlights the importance of incorporating prior systematics knowledge during training for successful denoising. Our work provides critical guidelines for using DL for 21cm foreground removal, tailored to specific data attributes. Notably, it is the first time that DL has been applied to the Planck sky model being most realistic foregrounds at present. 
 
\end{abstract}

\begin{keywords}
 -- methods: data analysis, statistical --radio continuum: general, galaxies --Cosmology: observations, large-scale structure of Universe 
\end{keywords}



\section{Introduction}

In the last few years, the 21cm intensity mapping (IM) has emerged as a new and promising technique to  probe the Epoch of Reionisation (EoR), study the large-scale-structures (LSS), and reveal the cosmic evolution history. This is achieved through the measurement of the neutral hydrogen (HI) fluctuations  as a tracer of matter distributions  in the Universe over  different redshifts. The concept of 21cm IM is to use a radio telescope with a relatively limited  resolution to map the HI line from multiple unresolved galaxies, tracing the LSS through intensity fluctuations over cosmological distances \citep{bdw04, pbp06}. The observing frequencies of the radio telescope provide accurate resolution in redshift, mapping the Universe through  three dimensional tomography.  This is complementary to the traditional approach of measuring LSS through an optical telescope, which detects individual galaxies with either photometric or spectroscopic redshifts \citep[e.g.,][]{aab+14, des17}.    

A number of HI IM experiments are currently operating or under development. Some experiments are designed as single dish telescope or use different dishes only in auto-correlation, such as GBT \citep{cpb+10, msb+13, wpm+21}, MeerKLASS \citep{sch+17, cls+22}, and BINGO \citep{bbd+13, afl+21}.  Others are designed as  interferometers, such as the CHIME telescope consisting of 4 cylinders with 256 antennas on each  \citep{chimeoverview, chimedetection}, and the HIRAX telescope constituting 1024 dishes \citep{nbb+16, hiraxoverview}.  The upcoming ska project will be possible  to conduct IM survey in single-dish mode at $0.35<z<3$ with SKA-MID, and in interferometer mode at $3<z<5$ with SKA-LOW \citep{skaredbook18}.

The detection of the HI signal is limited by the astrophysical foregrounds, which can be $\sim10^5$ times stronger than the HI signal and must be effectively removed in order to measure the HI signal \citep[e.g.,][]{ sss+15, ord16}. The frequency spectra of the foregrounds are smooth, while the HI signal is uncorrelated fluctuations independent of frequency. One relies on these different spectral characteristics to separate  the foregrounds from the HI signal \citep[e.g.,][]{msb+13, abf+15}. In addition, instrumental systematics, such as gain fluctuations \citep[e.g.,][]{wmc+22}, beam effects \citep[e.g.,][]{mss+21}, $1/f$ noise \citep[e.g.,][]{hdb+18, cbc+20}, polarisation leakage \citep[e.g.,][]{cic+21} and radio frequency interference \citep[RFI, e.g.,][]{hd18},  are another challenge which not only obscure the HI signal but also complicate the smoothness of the foreground spectral structure, impacting the effective component separation.

Many methods and algorithms have been developed to subtract the foregrounds from HI signals. Some are parametric or linear methods, which assume a physical model for sky components based on their known  properties. Examples of such parametric method include the Karhunen-Loeve Decomposition \citep{ssp+14} or the delay filter \citep{ekd+20}.  The limitation of such  methods is that they assume well-known instrumental response, and otherwise result in signal contamination due to  foreground residuals \citep[e.g.,][]{scc+21}. Non-parametric or blind methods, on the other hand, perform component separation based on statistical properties of the data without presuming a particular model.  Such methods include but not limited to principal component analysis (PCA) \citep[e.g.,][]{msb+13, smb+13, abf+15}, Independent component analysis (ICA) \citep[e.g.,][]{ abf+15, wba+16}, Generalised Morphological Componenet Analysis (GMCA) \citep[e.g.,][]{cib20}, and generalized needlet internal linear combination (GNILC) \citep[e.g.,][]{rdc11,ord16}. Non-parametric methods, however, are subject to over-subtraction and signal loss as their main disadvantage. 

The foreground subtraction for 21cm IM is similar to the image denoising problem in the deep learning (DL) field, where one aims to remove noise from an image. Indeed, a number of studies have proposed machine learning techniques for 21cm foreground removal. For example, \cite{lxm+19} showed that a convolutional denoising autoencoder is able to reconstruct EoR signal complicated with interferometric beam effects. \cite{mlv+21} demonstrated that a deep convolutional neural network (CNN) with a U-Net architecture can effectively separate foreground from 21cm single dish data. \cite{nlg+22} adopted the U-Net from \cite{mlv+21} and found that it can eliminate MeerKLASS-like primary beam effects better than PCA for foreground removal.  Following \cite{nlg+22} , \cite{gln+22} further showed that the U-Net based foreground subtraction is able to eliminate polarisation leakage. In recent studies, \cite{bgi+21, bgp+23} developed a U-Net based network to successfully identify neutral and ionized regions in 21cm EoR signal contaminated with foregrounds.

At the moment, all the above DL studies are based on simulated 21cm data subject to particular simulation models. Before widely applying deep neural networks to upcoming real observations, one needs to be cautious about their limitations and reliability.  Some key concerns include the consistency of performance across different sky models, the potential disparity between the training dataset and real observations, and the presence of systematics in the data. While there is still undergoing effort to generate more realistic simulations, it is useful to understand if DL is able to produce consistent results across different simulation models like other traditional approach such as PCA. Compared with previous studies,  we conduct a broad scope study on  the robustness of deep network under different  simulation approaches. In addition, we simulate the scenario of frequency-dependent instrumental systematics and investigate the network training strategy in such cases.

The paper is organised as follows. Section\,\ref{secunet} explains the architecture of our deep neural network.  Sections\,\ref{skysec}\,\&\,\ref{syssec} introduce our simulation models of the sky components and instrumental effects. Section\,\ref{anasec} describes the details of data processing for foreground removal. Section\,\ref{ressec} presents the results for different sky models and systematics contaminated data. 


\section{The U-Net} \label{secunet}
Our network is based on the U-Net model used in \cite{mlv+21}, which was firstly introduced by \cite{rfb15}. The U-Net is one type of convolutional neural network (CNN) which consists of an encoding path that captures input into representative features, and a decoding path that reconstructs the output from extracted features.  The encoding and decoding paths are connected by skip connections to allow the decoding to learn from lower-level input features.  Through a series of convolution operations (layers), the U-Net learns the important features from the provided input  data, and returns output after removing unwanted features to match the provided ground truth. 

Compared with other type of neural networks, a U-Net architecture is particularly suitable for foreground subtraction and the denoising problem in general. Its convolutional layers not only capture information at the single-pixel level but also consider the correlations between neighbouring pixels. This allows U-Net to represent both small  and large scale features of the data \citep{zzc+17}. The skip connections preserve low level features during denoising to  keep fine details \citep{msy16}. In addition, the flexible architecture of U-Net allows relatively easy modifications and adaptations for specific data and problems.  For our specific foreground subtraction task, the input data consist of HI signals and foregrounds. Ideally, the output from the network contains pure HI signals. Through the network training process, the U-Net aims to keep HI signals as the relevant feature and remove foregrounds as the unwanted feature.

Our network is created using  the \textsc{Keras} library \footnote{\url{https://keras.io}}. A table summarising our U-Net architecture is given in table\,\ref{tab:architect} in Appendix\,\ref{app:unet}. The encoding path of our U-Net has  four levels. Each level starts with three  convolution blocks. Each block consists of a 3D convolutional layer with a  $3\times3\times3$ kernel size, a normalisation layer that normalises the previous layer output over the batch sample to avoid over-fitting \citep{is15}, and an activation layer with a rectified linear unit (ReLU) activator for output \citep[e.g.,][]{gbb11}. We sets the padding option to be ``same'' for each convolutional layer, which means the data will be padded with 0 at the edges to preserve the same dimension as the layer input.  The first convolution block of each level doubles the feature channels at each level through its convolutional layer. The initial number of feature channels for the first level is chosen to be 32 to accommodate the GPU memory. The three convolution blocks are followed by a max-pooling layer \citep{jkr+09} to half the dimension of  data and  a dropout layer with a $20\%$ dropout rate to avoid over-fitting \citep{hsk+12, shk+14} before passing to the next level. The total number of network trainable parameters (weights) in our architecture is $\approx$34 million. We list the hyperparameters used for the network architecture in Table\,\ref{tab:unet}.
\begin{table}
\centering
\begin{tabular}{l|l}
\hline
\hline
Kernel size & $3\times3\times3$\\
Activation function & ReLU \\
Padding & Same \\
Initial No. of feature channels & 32\\
Dropout rate & $20\%$ \\
Stride  &  1 (encoder); 2 (decoder) \\
Loss function & Logcosh \\
Optimiser & NAdam \\
Learning rate & $10e^{-4}$ \\
Batch size & 16 \\
Total No. of trainable parameters & $\approx$34 million \\
\hline
\hline
\end{tabular}
\caption{The  hyperparameters for our U-Net architecture design and network training. The U-Net is created using the \textsc{Keras} library. }
\label{tab:unet}
\end{table}

The decoder path of our network is symmetric to the encoder path with five decoder levels. Each decoder level consists of three  transposed convolution blocks. Each block has a transposed 3D convolutional layer, a normalisation layer and an activation layer. The transposed 3D convolutional layer has an opposite scaling effect on the data dimension and  number of feature channels.  The first transposed convolutional layer  in each block halves the feature channels and doubles the dimension of data through a stride of 2.  The stride defines the step size of the convolution kernel moving across the data plane. The output from the  three transposed convolutional blocks is concatenated with the corresponding encoder level to preserve the fine features. A $20\%$ dropout layer is then applied before passing to the next decoder level. The final output consists of a 3D convolutional layer stacking all features  together.


\section{Sky models}\label{skysec}

We adopted three  sky models in our study to test the robustness of the DL network under different simulations. Our baseline results are based on the HI model in \cite{bbd+13} and foreground models in \cite{sck05} (hereafter MS model), where each component is simulated as a Gaussian realisation subject to its power spectrum. We further used the \textsc{CoLoRe} code to generate  HI realisations different than the MS model, where the structure formation is modeled by  the 1st-order Lagrangian perturbation theory \citep{rsa+22}.  In addition, we used the Planck sky model (PSM) to generate more realistic  foregrounds based on previously observed foreground templates \citep{dbm+13}. 

\subsection{MS model}

Our baseline results used the MS model for foreground simulation. We include Galactic synchrotron, Galactic free-free, extra-galactic free-free and extra-galactic point sources as our foreground emission. In the MS model, each  of these foreground component is simulated as a Gaussian field respecting a power-law spectrum in both angular, $\ell$, and frequency, $\nu$, space.  The power spectrum of these foreground components is modeled as \citep{sck05}
\begin{equation}\label{equ:ms}
C_{\ell} = A\left(\frac{1000}{\ell}\right)^{\beta}\left(\frac{\nu_{\rm ref}}{\nu}\right)^{\alpha}\,.
\end{equation}
The values of the  amplitude $A$, scale parameters $\beta$ and spectral index $\alpha$ for each component at a reference frequency $\nu_{\rm ref} = 130$\,MHz is listed in Table\,\ref{tabms}. The foreground maps are simulated at the reference frequency and scaled to other desired frequency channels with respect to the spectral index of each component. 
\begin{table}
\centering
\begin{tabular}{l|l|l|l}
\hline
\hline
& $A\,[\rm mK^2]$ & $\beta$ & $\alpha$ \\
\hline
Galactic synchrotron & 700 & 2.4 & 2.80 \\
Galactic free-free & 0.088 & 3.0 & 2.15 \\
Extra-galactic free-free & 0.014 & 1.0 & 2.10 \\
Point sources & 57 & 1.1 & 2.07 \\
\hline
\hline
\end{tabular}
\caption{Values of foreground parameters in Equ.\,\ref{equ:ms} under the MS model. The values listed here are at a reference frequency of 130\,MHz. }
\label{tabms}
\end{table}

The HI signal in our baseline results is simulated using the \cite{bbd+13} model, where the mean HI brightness temperature at a given redshift $z$ is calculated as  
\begin{equation}
\bar{T}_{\rm HI}(z) = 44\,\mu\text{K}\left(\frac{\Omega_{\rm HI}(z)h}{2.45\times10^{-4}} \right)\frac{(1+z)^2}{E(z)}\,.
\end{equation}
where $h = H_0/100$\,km\,s$^{-1}$\,Mpc$^{-1}$  is the Hubble constant, $\Omega_{\rm HI}h = 2.45\times10^{-4}$ is assumed to be constant over redshift, and $E(z) = H(z)/H_0$. The power spectrum of the HI signal at a particular redshift bin of $z_{\rm min}<z<z_{\rm max}$ is
\begin{equation}
  C_{\ell} = \frac{H_0b_{\rm HI}^2}{c}\int \text{d}z E(z)\left[\frac{\bar{T}_{\rm HI}D(z)}{(z_{\rm max}-z_{\rm min})r(z)}\right]^2P_{\rm cdm}\left(\frac{\ell+0.5}{r_0}\right)\,,
\end{equation}
where we assume a constant HI bias of $b_{\rm HI} = 1$ for simplicity, $D(z)$ is the growth factor, $r(z)$ is the comoving distance, and $P_{\rm cdm}$ is the underlying dark matter power spectrum at the current day ($z=0$). The HI maps, $T_{\rm HI}(z, \mathbf{\hat{n}})$,  are simulated as independent realisations at each frequency channel assuming they are uncorrelated along the frequency direction at a typical bin width of $\sim1\,$MHz for IM experiments. All cosmological parameters are set to the values in the \emph{Planck} 2018 cosmology results \citep{planckVI18}. Note that although the HI model in our baseline simulation is based on \cite{bbd+13}, for simplicity, we refer to the baseline simulation model as MS model throughout the paper.

\subsection{CoLoRe model} \label{sec:color}
In order to test the robustness of the DL network against different HI models, we replace the HI model from the MS model by using the \textsc{CoLoRe} package \citep{rsa+22}. According to the input cosmological parameters, the \textsc{CoLoRe} code initializes a cosmological model assuming a flat $w$CDM Universe. The code then generates a Gaussian realisation of the linear matter density field at $z=0$. The code supports three structure formation models to transform the linear matter density to a physical non-linear matter overdensity, namely,  the lognormal field \citep{cj91}, the 1st- and 2nd-order Lagrangian perturbation theory (LPT) \citep{bcg+02}. In our case, we selected the 1st-order LPT option to transform  to a non-linear HI overdensity field $\delta_{\rm HI}$ at different redshifts assuming $b_{\rm HI} = 1$. Compared with the lognormal field, the LPT has the advantages of capturing non-linear structures at smaller scales and avoiding extreme fluctuations caused by large values in the initial realisation of the linear matter field \citep{rsa+22}. The 1st-order LPT is used here to preserve non-linear structures at small scales while minimising the computational cost compared to the 2nd-order LPT.  The HI brightness temperature maps are generated by associating the HI temperature and the HI density through
\begin{equation}
  T(z) = \bar{T}_{\rm HI}(z) (1+\delta_{\rm HI}(z, \bm{\hat{n}}))\,.
\end{equation}
We refer the reader to \cite{rsa+22} for the details of the CoLoRe simulation tool.

The method of simulating the HI signal in the \textsc{CoLoRe} model is completely different and independent from that of the MS model. The former is based on an LPT structure formation model with non-linear matter distribution, and the later is a pure Gaussian realisation subject to the HI angular power spectrum. The two models thus test the robustness of the deep network against independent HI simulations. When adopting the \textsc{CoLoRe} model,  we keep the same foreground components as in the MS model and only replace the HI simulation. 

\subsection{Planck sky model}\label{sec:psm}

We replace the foreground components in the MS model by the PSM to test the robustness of the DL network against foreground models.  The PSM include Galactic synchrotron emission, free-free emission and extra-galactic point sources. The synchrotron emission is generated based on the reprocessed Haslam 408\,MHz all-sky Galactic synchrotron map \citep{hks+81, rdb+15}. The synchrotron map at each frequency channel is referenced to 408\,MHz through
\begin{equation}
  T(\nu, \hat{\bm n}) = T_{\rm 408}(\nu, \hat{\bm n})\left( \frac{\nu}{408\,\rm MHz}\right)^{\beta(\hat{\bm n})}\,,
\end{equation}
where we use \cite{pbb+98} for  the spatially-varying spectral index map $\beta(\hat{\bm n})$.

The optical $H_{\alpha}$ emission is a good tracer of free-free emission at radio frequencies.  We thus use the $H_{\alpha}$ map from \cite{ddd03} to generate  the free-free emission based on the $H_{\alpha}$-to-radio relation
\begin{equation}
  T \approx 10\,\text{mK}\left(\frac{T_e}{10^4\,\text{K}}\right)^{0.667}\left(\frac{\nu}{\text{GHz}}\right)^{-2.1}I_{H_{\alpha}}\,,
\end{equation}
where $T_e$ is the electron temperature at 7000\,K and  $I_{H_{\alpha}}$ is the full-sky $H_{\alpha}$ emission map in Rayleigh. 

We follow \cite{ord16} to simulate extra-galactic point sources, which has three components: i) a mean background temperature; ii) the Poisson distributed sources; iii) the clustered sources. The mean background temperature is computed by
\begin{equation}
\bar{T}_{\rm PS} \propto \int_0^{S_{\rm max}} S\frac{\text{d}N}{\text{d}S}\text{d}S\,,
\end{equation}
where $\frac{\text{d}N}{\text{d}S}$ is the source count computed from observed continuum data at 1.4\,GHz. $S_{\rm max} =  0.1$\,Jy in our case is the cut-off source flux, above which one can subtract the bright sources by cross-matching with past surveys. The Poisson distributed sources has two components. Firstly weak sources are simulated by  a Gaussian realisation with a power spectrum of
\begin{equation}
C_{\ell}^{\rm Poisson}\propto\int_0^{S_{\rm ps}}S^2\frac{\text{d}N}{\text{d}S}\text{d}S\,,
\end{equation}
where $S_{\rm ps} = 0.01$\,Jy is the upper limit flux of sources that satisfy the Gaussian distribution. Secondly, for bright sources with $S_{\rm ps}<S <S_{\rm max}$, we randomly allocate their flux values and directly distribute them on the sky. Finally,  we simulate clustered sources with a power spectrum of
\begin{equation}
  C_{\ell}^{\rm Cluster} \propto \ell^{-1.2}\bar{T}_{\rm ps}^2\,.
\end{equation}
The point source maps of all added components above are scaled along the frequency direction with a mean spectral index of $-2.7$ and a standard deviation of 0.2 for different pixels \citep{scc+21}. 

We refer the reader to \cite{ord16} and \cite{bbd+13} for detailed description of the PSM. As the PSM is based on observed templates, it is completely different than the pure Gaussian realisations of the MS foreground model. By keeping HI simulations the same, we use the PSM and the MS model to test the robustness of the deep network against different foreground models. One limitation of the PSM model is that  it is non-trivial to simulate different sky realisations for network training or testing purposes, since it is modeled  based on the true sky. To vary foregrounds in this case, we replace the spectral index map of the synchrotron emission from \cite{pbb+98}  by normal randomisations with a mean of -2.9 and a standard deviation of 0.5. Since the synchrotron emission is  dominant over other components in the data, we find that by varying the synchrotron spectral index alone, it changes the sky enough to be considered as different realisations.  

\section{Instrumental effects} \label{syssec}
In this section, we describe the survey and instrumental parameters assuming a SKA-MID Band\,1-like experiment. In addition, to test the robustness of the deep network against  frequency structures, we apply a frequency-dependent Gaussian beam \ref{sec:beam} and a sinusoidal gain drift \ref{sec:gain} to the  simulated data under the baseline MS model. 

\subsection{Survey parameters}\label{secsuv}

\begin{table}
\centering
\begin{tabular}{l|l}
\hline
\hline
Dish diameter, $D$ (m) & 15 \\
Frequency range, $\Delta\nu$ (MHz) & [700, 1020]\\
No. channels, $N_{\rm bin}$ & 64 \\
Channel width, $\delta\nu$ (MHz) & 5 \\
Redshift range, [$z_{\rm min}$, $z_{\rm max}$] & [0.4, 1.0]\\
Beam resolution, [$\theta_{\rm min}$, $\theta_{\rm max}$] (deg) & [1.1, 1.6] \\
\hline
\hline
\end{tabular}
\caption{Survey and instrumental parameters used in our simulation. The simulation is a sub frequency band of the SKA-MID Band\,1 \protect\citep{skaredbook18}. }
\label{tab:ska}
\end{table}

One of the motivations of applying DL in the astronomy community is that DL is able to handle massive datasets from upcoming large surveys such as the SKA. The first phase of SKA consists of two telescopes - SKA-LOW and SKA-MID. SKA-LOW observes at $3<z<6$ with the main goal of studying the epoch of reionisation (EoR). SKA-MID focuses on the low redshift range of $z<3$ for studying large-scale-structures (LSS) and Dark Energy. SKA-MID has two frequency bands. Band\,1 will observe at 350\,MHz$-$1050\,MHz, and Band\,2 will observe at 950\,MHz$-$1750\,MHz. For a 21cm survey,  SKA-LOW will operate in the interferometer mode while SKA-MID will operate in the single dish mode \citep{skaredbook18}.  

In this study, we evaluate  the stability of DL network on single dish observations with frequency distortions at low redshift.  For our simulation, we consider a sub frequency band of the SKA-MID Band\,1 between the frequency range of 700\,MHz and 1020\,MHz, corresponding to a redshift range of $0.4<z<1.0$. We divide the frequency band into 64 frequency channel with a channel width of 5\,MHz. The choice of the number of frequency channels is to facilitate the foreground removal. Too few channels will impair both the pre-processing PCA and the DL network for foreground removal (see Sect.\,\ref{anasec}). Too many channels will result in significant computational cost. The SKA-MID telescope has a dish diameter of 15\,. In single dish mode, it gives a beam resolution of $\sim$1.3 degree in our frequency range.

Table\,\ref{tab:ska} summarises the survey and  instrumental parameters used in our simulation. We do not include the instrumental thermal noise in our simulation as the purpose of our work is to test the network under different sky models and frequency-dependent systematics. The thermal noise is thus not critical for the specific scope of this work and in any case, one usually removes the thermal noise through cross-correlating data from different seasons \citep[e.g.,][]{wmc+22}.

\subsection{Beam model}\label{sec:beam}
We focus on studying the stability of the network against frequency-dependent instrumental effects, we thus do not consider complicated beam features but adopt a simple Gaussian beam model in our data. The full width at half maximum (FWHM) of the beam is calculated as
\begin{equation}\label{equ:beam}
  \theta = \frac{\lambda}{D}\,,
  \end{equation}
where $D$ is the telescope dish diameter and $\lambda$ is the observing wavelength. We simulate two scenarios. Firstly, the FWHM is fixed to the lowest resolution (highest observing frequency) for all frequency channels. This is a common approach to simplify  the frequency evolution of the beam size to facilitate the foreground removal \citep[e.g.,][]{smb+13}. Unless otherwise stated in the paper, we apply the frequency-fixed beam to all data by default. For example, when evaluating the performance of the U-Net under different sky models, the frequency-fixed beam is applied throughout for consistency.

The frequency-fixed beam, however, loses information by degrading the beam resolution at higher frequency channels. In the second scenario, we keep the frequency evolution of the beam size scaling as $\frac{\lambda(\nu)}{D}$. We test the robustness of the network against the frequency-dependent beam in Section\,\ref{secbeam}. 

\subsection{Gain drift model}\label{sec:gain}
In addition to beam effect, we introduce frequency-dependent gain drift  in our data to test the reliability of the DL network in the presence of gain fluctuations.  Gain drift is a systematic change in the gain response of a telescope. It can be caused by several factors such as the outside temperature change or calibration errors of the telescope. A frequency-dependent gain drift introduces additional irregularities in the frequency direction,  complicating the component separation algorithm which traditionally relies on the smoothness of foreground spectra.

The gain drift applies to the data as
\begin{equation}\label{equ:gain}
\tilde{d}(\nu, \bm{\hat{n}}) = G(\nu)d(\nu,\bm{\hat{n}})\,,
\end{equation}
where $\tilde{d}(\nu, \bm{\hat{n}})$  is the data contaminated by gain drift and $d(\nu,\bm{\hat{n}})$ is the simulated data from the sky. In the ideal scenario where no gain drift exists, one has $G(\nu) =1$. In our work, we adopt the model from \cite{mrc+16} to simulate a frequency-dependent gain drift. The gain drift  in this case has a sinusoidal pattern as a function of observing frequency such that 
\begin{equation}\label{equ:gv}
G(\nu) = G_0\sin (G_1\nu + G_2)+1\,.
\end{equation}
We set the mean amplitude parameter $\bar{G}_0 = 0.5$ so that the gain fluctuation varies between  0.5 and 1.5. We note that a gain fluctuation of 0.5 is relatively large for a radio telescope. We however keep $\bar{G}_0=0.5$ to test the network in the worst scenario. The mean period parameter $\bar{G}_1$ is set to be 1 to create enough cycle of oscillations  among the 64 frequency channels. We set the mean phase parameter $\bar{G}_2 = 0.5$ although the phase parameter has no effect on our main result. To  create variations in the gain drift among different realisations of simulated data for training, validation and testing,  the values of $G_0$, $G_1$ and $G_2$ are randomly sampled from a normal distribution  with a standard deviation of 0.01 for each realisation.

\section{Data analysis}\label{anasec}
The \textsc{Keras}-based U-Net requires the input 3D data to be a  cube in a cartesian grid.  In our case, the simulated maps subject to different models are firstly generated in the \textsc{Healpix} scheme \citep{ghb+05}. This is because both the \textsc{CoLoRe} simulation package and the PSM templates are in the \textsc{Healpix} scheme. The \textsc{Healpix} maps are then divided into 3D cubes with dimension of [$N_{\rm pix}$, $N_{\rm pix}$, $N_{\rm freq}$] for training, validating and testing the U-Net. We also apply PCA as a quick pre-processing step to reduce the dynamic range between foregrounds and the signal, which is critical for the U-Net to effectively recover the signal afterwards. We describes the details of  data processing in the following section. 

\subsection{Data assembly}
For each sky model, we create 40 \textsc{Healpix} realisations of training maps and 10 realisations of validation maps.  We choose \textsc{nside = 256} so that the pixel size ($\sim13.7$ arcmin) is below the instrumental beam size ($\sim1.3$ degree), and that the total number of pixels is not too large to be computationally expensive. Each sky realisation also has 64 frequency channels with 5\,MHz bandwidth, and is convolved with the beam resolution according to Section\,\ref{secsuv}.  For each \textsc{Healpix} map with \textsc{nside = 256}, we divide into 192 sky patches of 64$\times$64 pixels with equal sky area.  Since the U-Net reconstructs the signal on an image by image basis, each of the 192 sky patches can be treated as an independent training sample for the network. We thus have 7680 training samples from  40  \textsc{Healpix} realisations, and 1920 validation samples from 10  \textsc{Healpix} realisations. The final input and output data format for the U-Net is thus a 64\,$N_{\rm pix}\times$64\,$N_{\rm pix}\times$64\,$N_{\rm freq}$ cube. 

For testing, we simulate 10 \textsc{Healpix} maps (equivalently 1920 test samples) for the MS model and \textsc{CoLoRe} model respectively. For the PSM, the training and validating samples are generated by varying the spectral index map of the synchrotron emission. For testing under the PSM, we use the true \cite{pbb+98} spectral index map for the synchrotron emission, which only gives one  \textsc{Healpix} map, equivalent to 192 test samples. 

\subsection{Data pre-processing}
By itself, the U-Net is not able to handle the large dynamic range between HI signals and foregrounds. This has also been previously reported by, e.g., \cite{mlv+21}, \cite{nlg+22} and \cite{bgp+23}. Without additional pre-processing steps, the $\sim5$ order of magnitude difference between HI and foregrounds makes the U-Net extremely difficult to correctly detect and recover the HI fluctuations in the maps. Therefore, we pre-process the data before feeding into the network by applying a component separation algorithm  to pre-reduce the foreground level.

We use Principle Component Analysis (PCA) as our pre-processing algorithm since it can effectively reduce the strong foreground emission without assumptions of specific sky models. In addition, it is quick and easy to apply as a pre-processing step. Nevertheless, we expect no major change on our main conclusions with other pre-processing algorithms. The PCA performs a singular value decomposition and identifies the principle components with the most significant variances in the data \citep[e.g.,][]{scc+21}. We use the \textsc{sclearn} package\footnote{\url{https://scikit-learn.org/stable/modules/generated/sklearn.decomposition.PCA.html}}  to apply PCA along the frequency direction summing over all pixels, since the correlation along frequency is the main distinguishable feature  between foregrounds and the signal. The PCA is applied to \textsc{Healpix} maps before dividing the maps into sky patches so that the PCA induced artifacts are more stable among samples for the network. We remove the first two principal components  for pre-processing and feed the PCA pre-processed data into the U-Net for fine-tuning. We have tested that removing less than two PCA modes is not enough to reduce the dynamic range but removing more than two PCA modes will leave limited space for improvement with the U-Net and can result in signal loss. 


\begin{figure}
\includegraphics[width = 0.97\hsize]{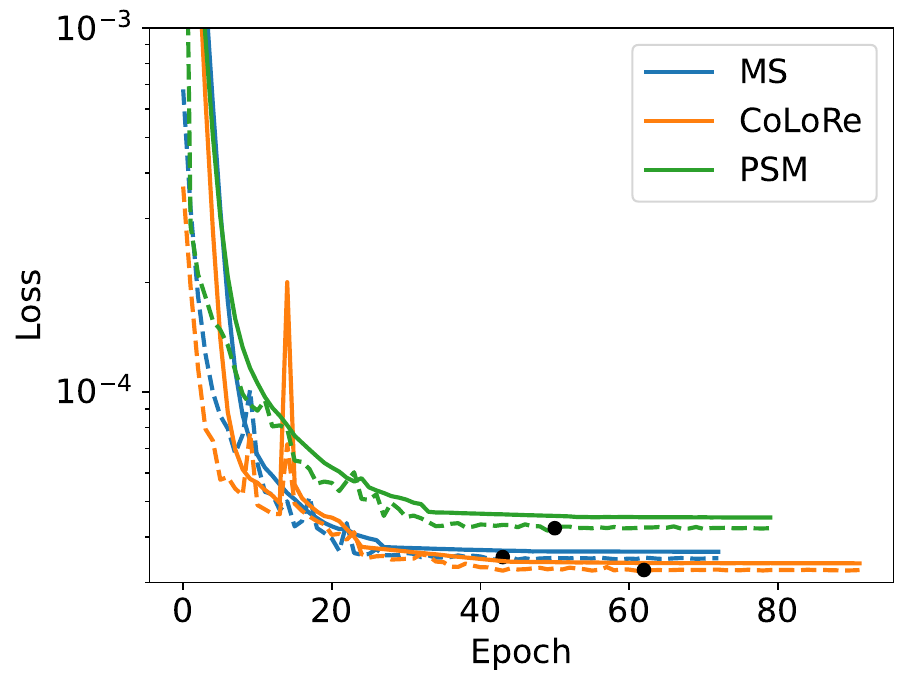}
\caption{The loss function as a function of training epoch for the three sky models. For each model, we show the training (\emph{solid}) loss and the validation (\emph{dashed})  loss.  The ending epoch in each case is reached when the validation loss is not improving for a continuous of 30 epochs. The optimal epochs corresponding to the least validation losses are indicated by the \emph{black dots}. }
\label{fig:loss}
\end{figure}

\subsection{Network training}
We train the network with a batch size of 16 limited by the GPU memory.  The U-Net output is evaluated by a  loss function that measures the difference between the network output and the true HI signal. During the training, the network trainable parameters (weights) are adjusted to by an optimizer to minimise the value of the loss function. We use Log-Cosh as our loss function as it is known to produce smooth image outputs \citep[e.g.,][]{mlv+21}. The Log-Cosh loss function is defined by
\begin{equation}
  \mathcal{L} = \sum_i \text{log cosh}(p_i-t_i)\,,
\end{equation}
where $p_i$ and $t_i$ are the output of the network and the true signal at each voxel.  The NAdam optimizer \citep{doz15} is used to adjust the network weights throughout the training. The optimal learning rate in our case is found to be $10e^{-4}$ initially and decays by a factor of 10 if the validation loss is not improving after 20 epochs. The network stops training if the validation loss is not improving after 30 epochs to avoid over-fitting.  The optimal epoch which gives the least validation loss value is saved to denoise the test dataset. Our choice of the network training hyperparameters are summarised in Table\,\ref{tab:unet}.

To evaluate the performance of the network under different sky models, we train and test the network with each model separately. Fig.\,\ref{fig:loss} shows the loss functions of the three different sky models.  The ending epochs and the optimal epochs (\emph{black dots}) vary among models. In all cases, the validation losses decrease steadily along epoch and stabilise at slightly lower values than the training losses after $\sim$50 epochs. The loss functions  indicate that the network is successfully trained to learn relevant features and accurately reconstructs signals from  unseen data. The PSM has the largest loss values because its observation-based foreground components are more complex than the other models. In particular, the strong Galactic emissions near the Galactic plane introduce large variations among training samples. It is thus more difficult for the network to learn the features and accurately reconstruct signals for the PSM. 



\begin{figure*}
\includegraphics[width = 0.97\hsize]{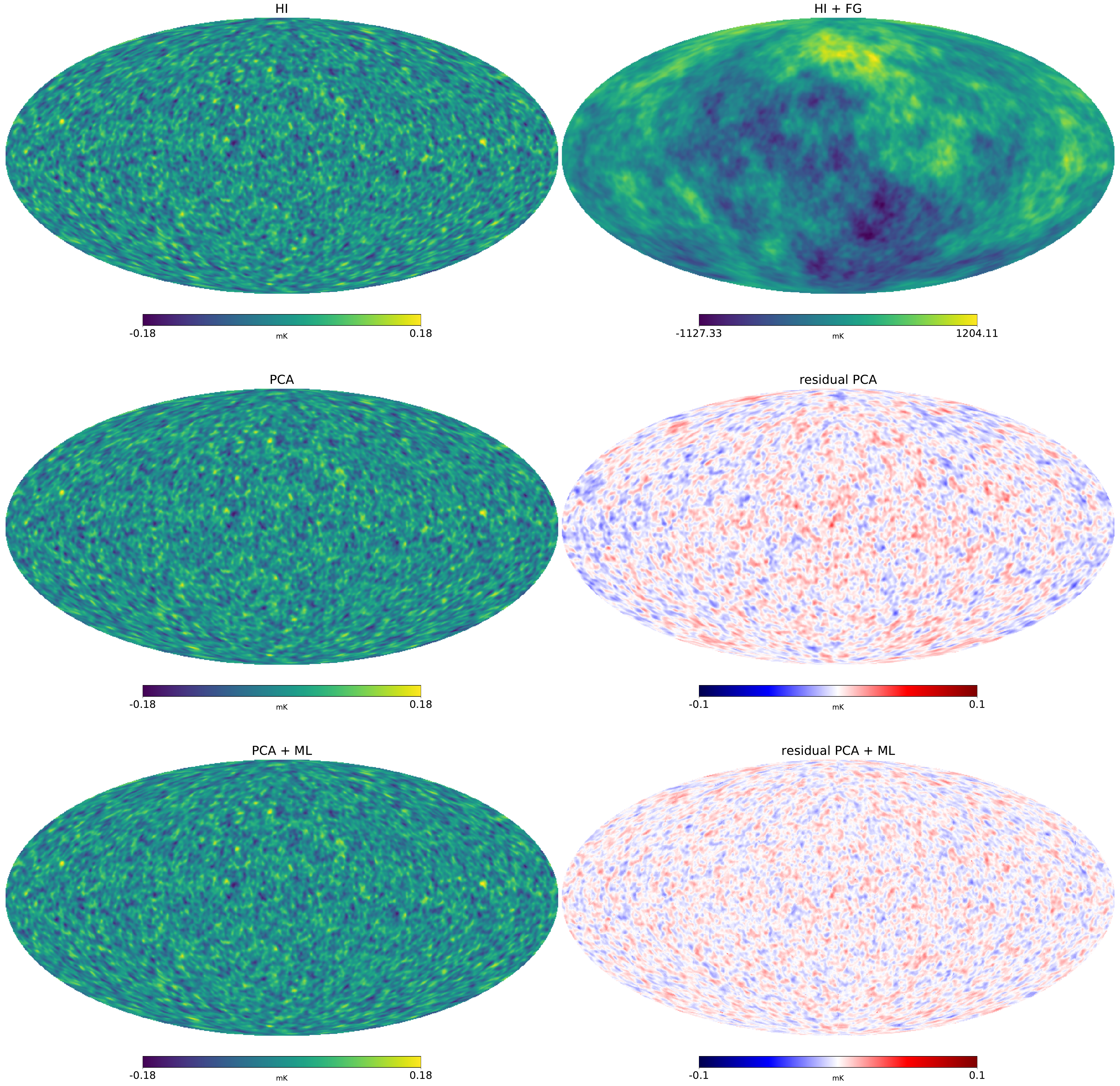}
\caption{The map results of foreground removal from the MS simulation model. The top panels show  the true HI map and the total HI + FG map. The middle panels show the reconstructed HI map from PCA\,2 mode  alone and its  residual. The bottom panels are  the reconstructed HI map from PCA\,2 mode + ML and its residual.  The plots are from a single realisation of \textsc{healpix} maps in the test dataset, centred at a single frequency channel at 862.5\,MHz ($z \approx$0.65). The reconstructed maps from the network  are reassembled back to a full sky \textsc{healpix} map from  individual sky patches. }
\label{fig:msmap}
\end{figure*}

\begin{figure}
\includegraphics[width = 0.97\hsize]{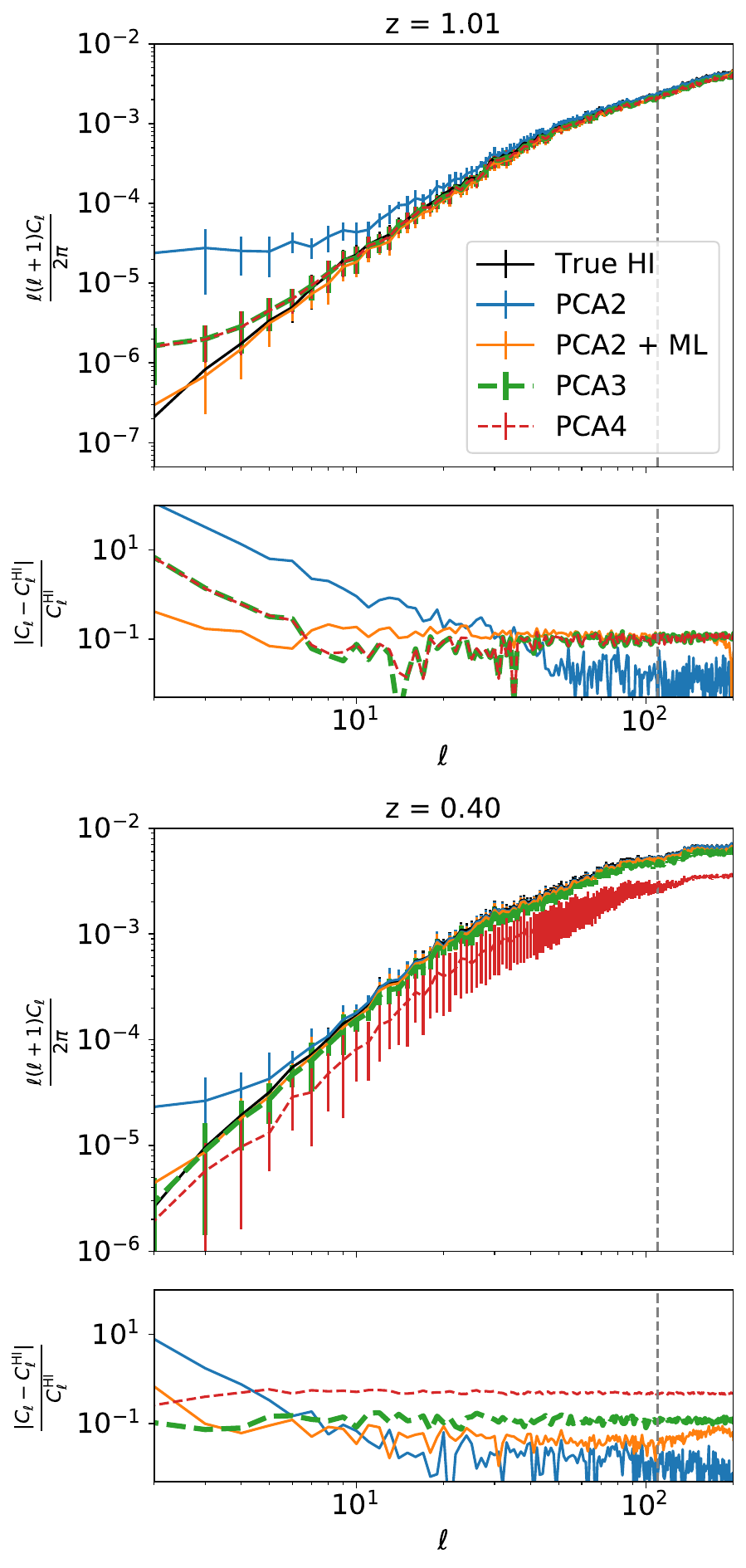}
\caption{The angular power spectra of the reconstructed signal and their fractional residuals under the MS simulation model. The power spectra at a higher redshift of $z=1.01$ (\emph{upper})and a lower redshift of $z = 0.40$ (\emph{lower}) are shown as an example.  In each redshift, we compare the true HI map (\emph{black}) with  the reconstructed HI maps from PCA 2 alone (\emph{blue}), PCA 2 + ML (\emph{yellow}), and PCA alone with 3 and 4 mode removal respectively (\emph{dashed}). The fractional residuals from the  reconstructed spectra with respect to the true HI spectrum are plotted in the lower sub-panel for each redshift.   The vertical grey dashed line indicates the corresponding angular scale due to beam resolution.  All power spectra are computed from the \textsc{healpix} whole sky maps after 're-assembling' the sky patches  output by the network, and have been beam-corrected. The plotted spectra are the means of the 10 \textsc{healpix} test samples with the error bars calculated from the standard deviation.   }
\label{fig:mscl}
\end{figure}
\section{Results}\label{ressec}
We first show  the results of the U-Net under different sky simulations in section\,\ref{res:sky}. We evaluate the network performance through map visualisation, power spectrum calculation and $R^2$ score comparison. In each case, the network is trained, validated and tested on the same sky model. We also present a more challenging scenario where  the training and testing dataset are from different models. In all cases, the simulated  data are convolved with the frequency-fixed beam as described in Section\,\ref{sec:beam}. The MS sky model is used as our baseline model since it simulates each sky component as a Gaussian field without specific assumptions and it is relatively fast to generate many data samples. In section\,\ref{res:sys}, we present the network reconstruction results when the data are convolved with frequency-dependent beam or contaminated with frequency-dependent gain drifts.

\begin{figure*}
\includegraphics[width = 0.322\hsize]{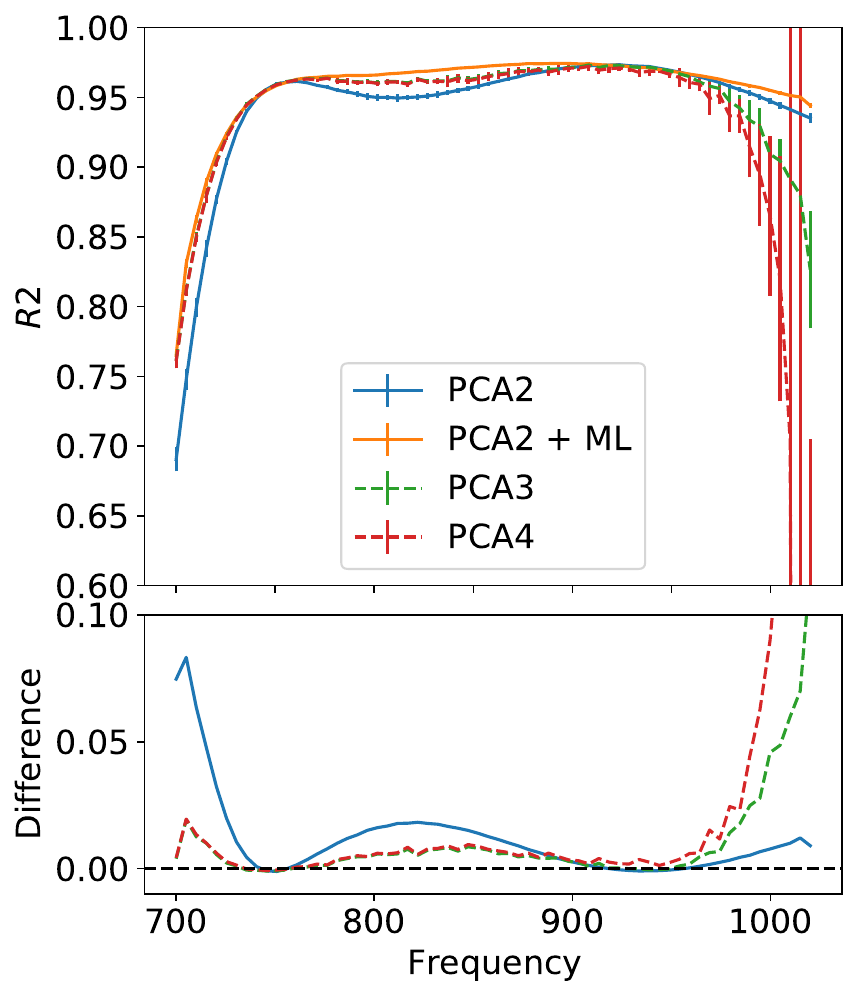}
\includegraphics[width = 0.315\hsize]{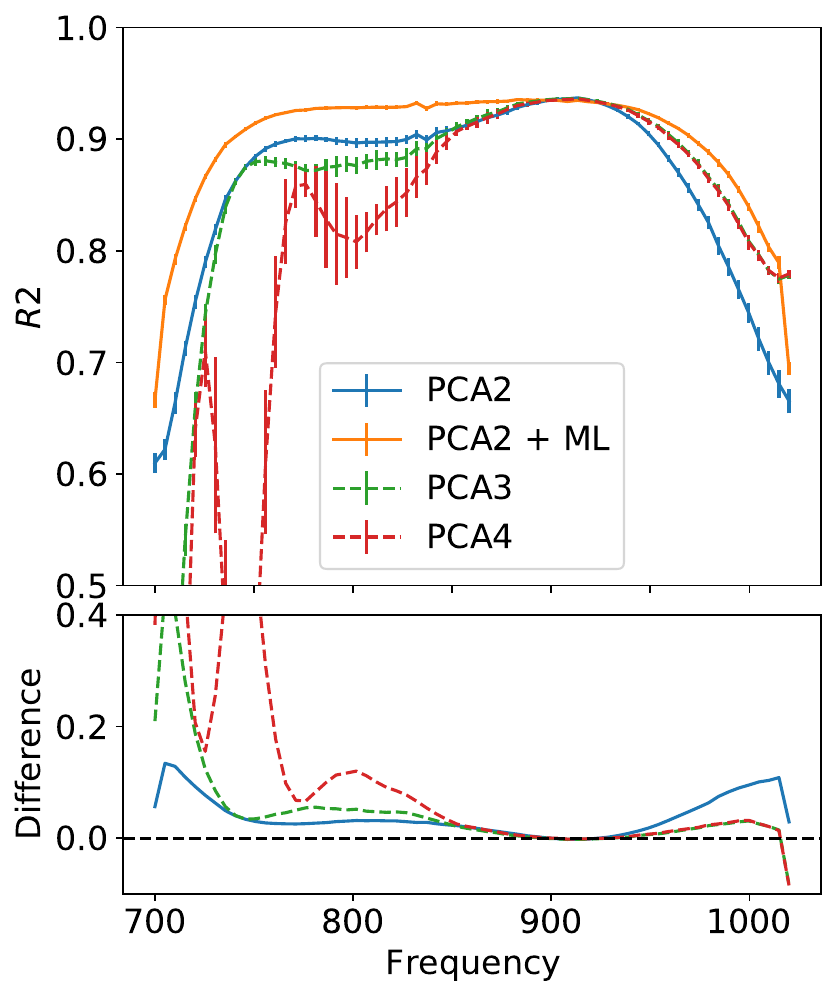}
\includegraphics[width = 0.315\hsize]{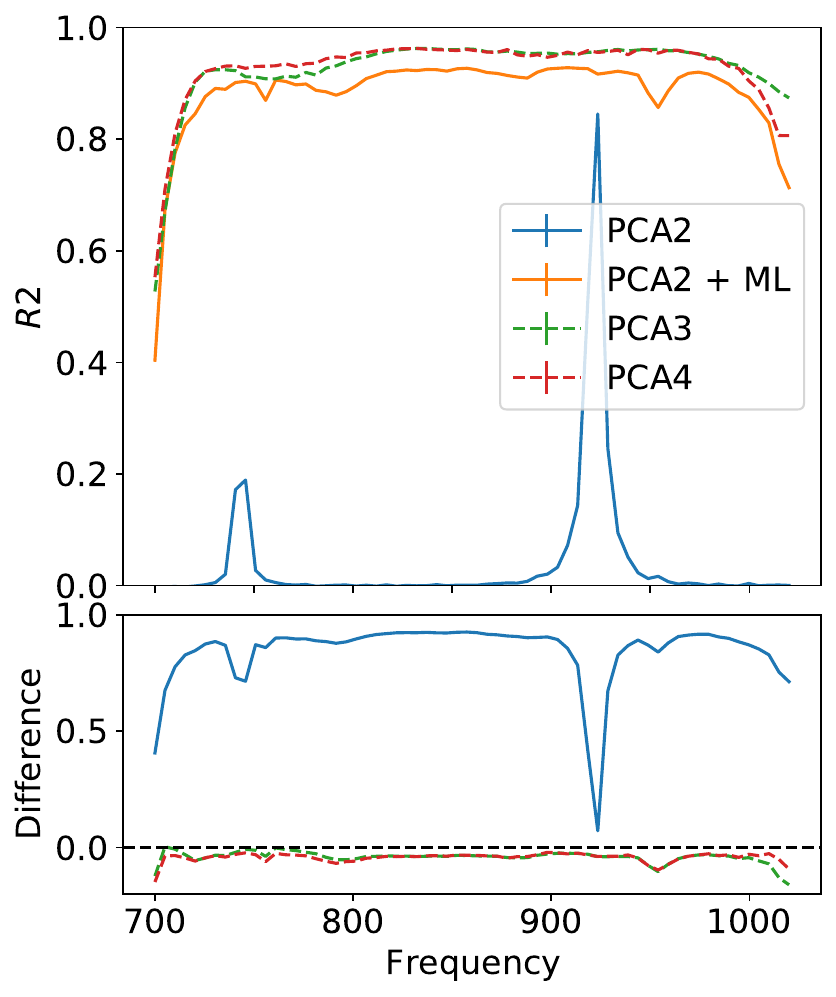}
\caption{The coefficients of determination ($R^2$ score) of the reconstructed signal from the MS simulation model (\emph{left}), the \textsc{CoLoRe} simulation model (\emph{middle}) and the Planck sky model (PSM, \emph{right}). In each case,  we compare the $R^2$ score from the pre-processing PCA 2 alone  (\emph{blue}),  PCA2 + ML (\emph{yellow}), and PCA 3 and  4 mode removal (\emph{dashed}).  The bottom sub-plots show the differences  between the U-Net reconstructed $R^2$ and those from PCA alone, with the color coding corresponding to the upper sub-plots. All $R^2$ scores are computed from the \textsc{healpix} whole sky maps after 're-assembling' the sky patches, and are computed as a  function of  the observing frequency. For the MS and \textsc{CoLoRe} simulation models, we show the  mean $R^2$ score of the 10 \textsc{healpix} test samples  with error bars calculated from the standard deviation. For the PSM model, the $R^2$ score is based on the single  \textsc{healpix}  test sample  as the PSM model provides only one realisation of the foregrounds. }
\label{fig:modelr2}
\end{figure*}

\subsection{Sky model dependency}\label{res:sky}
\subsubsection{MS model}\label{sec:ms}
We create baseline results using the MS sky model. In this case, the network is trained, validated and tested consistently under different realisations of the MS model. Fig.\,\ref{fig:msmap} shows the full sky maps of the input data and the reconstructed signals from PCA two mode removal alone (PCA\,2) and from the U-Net in addition to the PCA\,2 pre-processing (PCA\,2 + ML). The maps are reassembled back to \textsc{healpix} schemes from $64\times64\times64$ cubes output by the network. Here we show one  \textsc{healpix} realisation from the test dataset at a medium frequency channel of 862.5\,MHz as an example. Compared with the original input data with foregrounds (\emph{top right}), both PCA\,2  (\emph{middle left}) and PCA\,2 + ML (\emph{bottom left}) significantly reduce the foregrounds. Both reconstructed maps are comparable with the true signal (\emph{top left}) in terms of the amplitude and features of fluctuations. Their residuals can be computed by subtracting the true HI map from the reconstructed maps respectively.  By comparing  the amplitude of the residual maps, the PCA\,2 + ML residuals (\emph{bottom right}) are lower than that of  the PCA\,2  residuals (\emph{middle right}). The map results provide a first indication that under the MS sky model, the output from the network is indeed matching the true signal more closely with less residuals.

To quantitatively compare the results from the network with those from  PCA alone, we compute the angular power spectra of reconstructed signals. Fig.\,\ref{fig:mscl} shows the power spectra of two frequency bins at the higher and lower ends of our  redshift coverage for comparison. The power spectra are computed from reassembled \textsc{healpix} maps with the beam effect corrected using the \textsc{anafast} function of the \textsc{healpy} package.  In each case, we plot the averaged power spectrum value per multipole over all 10 \textsc{healpix} maps in  the test dataset along with its standard deviation.  In each redshift, we compare the power spectrum from PCA\,2 + ML with PCA\,2 alone and the true HI. In addition, we also show the results from PCA alone removing  three (PCA\,3) and four (PCA\,4) modes. This is to understand whether the network gives consistent or better results than simply removing more modes with PCA.

From Fig.\,\ref{fig:mscl}, at the higher redshift bin of $z = 1.01$, all of the PCA alone power spectra show an excess of power at large scales below $\ell\sim10$ compared with the true HI spectrum (\emph{black}). This shows the presence of foreground residuals in the reconstructed signal after simple PCA modes removal, since foregrounds dominate at large scales. The U-Net spectrum (\emph{orange}), in comparison, fits best the true spectrum. In the lower sub panel, we plot the fractional residual defined as
\begin{equation}
\delta = \frac{|C_{\ell}-C_{\ell}^{\rm HI}|}{C_{\ell}^{\rm HI}}\,,
\end{equation}
where $C_{\ell}$ is  the reconstructed spectrum  from each method and $C_{\ell}^{\rm HI}$ is the true HI spectrum. The quantity $\delta$ gives the absolute residual in the reconstructed spectra with respect to the signal. By comparing the fractional residual curves at $z = 1.01$ in Fig.\,\ref{fig:mscl}, the PCA alone results indeed leave more residuals compared with PCA\,2 + ML at large scales due to leftover foregrounds. The PCA\,2 + ML  (\emph{orange})  has an almost scale-dependent fractional residual constant at  $\delta\sim10\%$ of the signal.  The network  surpasses the PCA alone performances at large scales of $\ell\lesssim10$ in terms of reducing foreground residuals. Meanwhile,  the  network does not show more signal loss than PCA alone  at smaller scales of $\ell\gtrsim10$ as its fractional curve is consistent with PCA\,3 and PCA\,4.  We note that the PCA\,2 alone method (\emph{blue}) has less residuals at $\ell \gtrsim50$ than other methods. This is because by removing less PCA mode, one reduces the over-subtraction of the signal which dominates primarily at smaller scales. However, this is at the cost of having the most foreground residuals at large scales and thus PCA\,2 alone is hardly the most interesting approach in this case.

At the lower redshift bin of $z = 0.40$ in Fig.\,\ref{fig:mscl}, the network performance is comparable with PCA\,3. In the upper subpanel, both the PCA\,2 + ML  (\emph{orange}) and the PCA\,3 spectra (\emph{green}) closely fit the true spectrum (\emph{black}). The PCA\,2 spectrum (\emph{blue}) shows excessive foreground at large scales and PCA\,4 over-subtracts the signal resulting in  the biased spectrum with low power (\emph{red}).  This result  is confirmed in the fractional residual curves in the lower sub-panel, where PCA\,2 + ML and PCA\,3 residuals are comparable. However, the PCA\,2 + ML shows slightly lower fractional residuals than PCA\,3 at small scales of $\ell\gtrsim10$, indicating less signal loss at these scales. In comparison, the PCA\,2 show much higher residuals at large scale and PCA\,4 observes large  fractional residuals at all scales.  By comparing the two redshift cases in Fig.\,\ref{fig:mscl}, the PCA alone results are more sensitive to the redshift. For example, PCA\,4 does not subtract enough foreground at $z = 1.01$ but over-subtracts the signal at $z = 0.40$. This is because the HI signal is weaker at higher redshift and thus it is more challenging for PCA to completely subtract foregrounds. At lower redshift, the HI signal is stronger and thus tends to suffer from signal loss due to the same number of mode removal as in the higher redshift.  On the contrary,  the performance of the network is less sensitive to the redshift, with an average fractional residual of $~10\%$ of the signal over all scales in both cases. This is potentially because the U-Net subtracts foregrounds by not only relying on the smooth foreground frequency response, but also by  matching the signal in the image space. Therefore, it is less affected by the change of foreground to signal ratio at different redshifts.

We also compute the coefficient of determination, denoted $R^2$ score,  as an additional measurement to evaluate the performance of the U-Net  on reconstructing signals, defined as
\begin{equation}
R^2 = 1 - \frac{\sum_i (t_i - o_i)^2}{\sum_i (t_i - \bar{t})^2}\,,
\end{equation}
where $t_i$ and $o_i$ are the true  and output values from the network of a data point $i$ in a dataset. $\bar{t}$ is the mean of the true data. The $R^2$ score quantifies how well the output from a model fits  the truth. In the ideal case, the model perfectly predicts the variance of the true data, and one obtains $R^2 = 1$. In general, one has $R^2<1$, meaning that the model partially accounts for the variance in the data, with a higher score indicating a closer match. A value of $R^2 = 0$ means that the model's prediction is no better than just the mean of the true data. In such case, we compute the $R^2$ score on the reassembled \textsc{healpix} maps of our test dataset from each method with respect to the true HI map. Each data point $i$ resembles a pixel in the \textsc{healpix} maps. Our computed $R^2$ scores  evaluate each method in the image space based on the output maps. 

The left panel in Fig.\,\ref{fig:modelr2} shows the $R^2$ scores of  reconstructed signals under the MS model. The $R^2$ score of each method is calculated per frequency channel by averaging over the $R^2$ scores of the 10 output \textsc{healpix} maps from the test dataset. In the upper subpanel, we show the mean $R^2$ score of each method as a function of frequency as well as its standard deviation. In the lower subpanel, we show the difference in $R^2$ scores between PCA\,2 + ML and PCA alone by subtracting the $R^2$ values of each PCA alone method respectively from that  of PCA\,2 + ML. A positive value in the $R^2$ difference means a worse reconstruction of the signal from PCA alone. We can see from the upper subpanel that PCA\,2 + ML has the highest $R^2$ scores at all frequencies. This is highlighted in the lower subpanel that the $R^2$ differences are positive for all PCA alone methods at all frequencies. The $R^2$ score results demonstrate that the network surpasses the PCA alone methods in terms of matching the output maps towards the true HI map. This is because the U-Net is trained to match the true HI in the 3D dimension through convolutional layers, and thus has advantages over the PCA alone methods when evaluated in the image space.  The drop of $R^2$ values at the  two frequency ends  is due to the lack of continuous  frequency information beyond the two ends. Since the PCA is applied along the frequency direction, the frequency discontinuation deteriorates the performance of PCA at the two ends, which also impacts the network efficiency due to pre-processing.

Overall, we have seen that under the MS sky model with our particular dataset, the network performance  is comparable with PCA for foreground removal with some advantages at certain cases. The U-Net is less sensitive to the redshift, and is more efficient on removing large scale foreground emission particularly at higher redshift than PCA. Based on the power spectrum reconstruction, the network is comparable with PCA\,3  alone but has less  foreground residuals at higher redshift and less signal loss at lower redshift. On the map level, the U-Net is more accurate in matching its output towards the true HI maps thanks to its 3D convolution operations. 

\begin{figure*}
\includegraphics[width = 0.97\hsize]{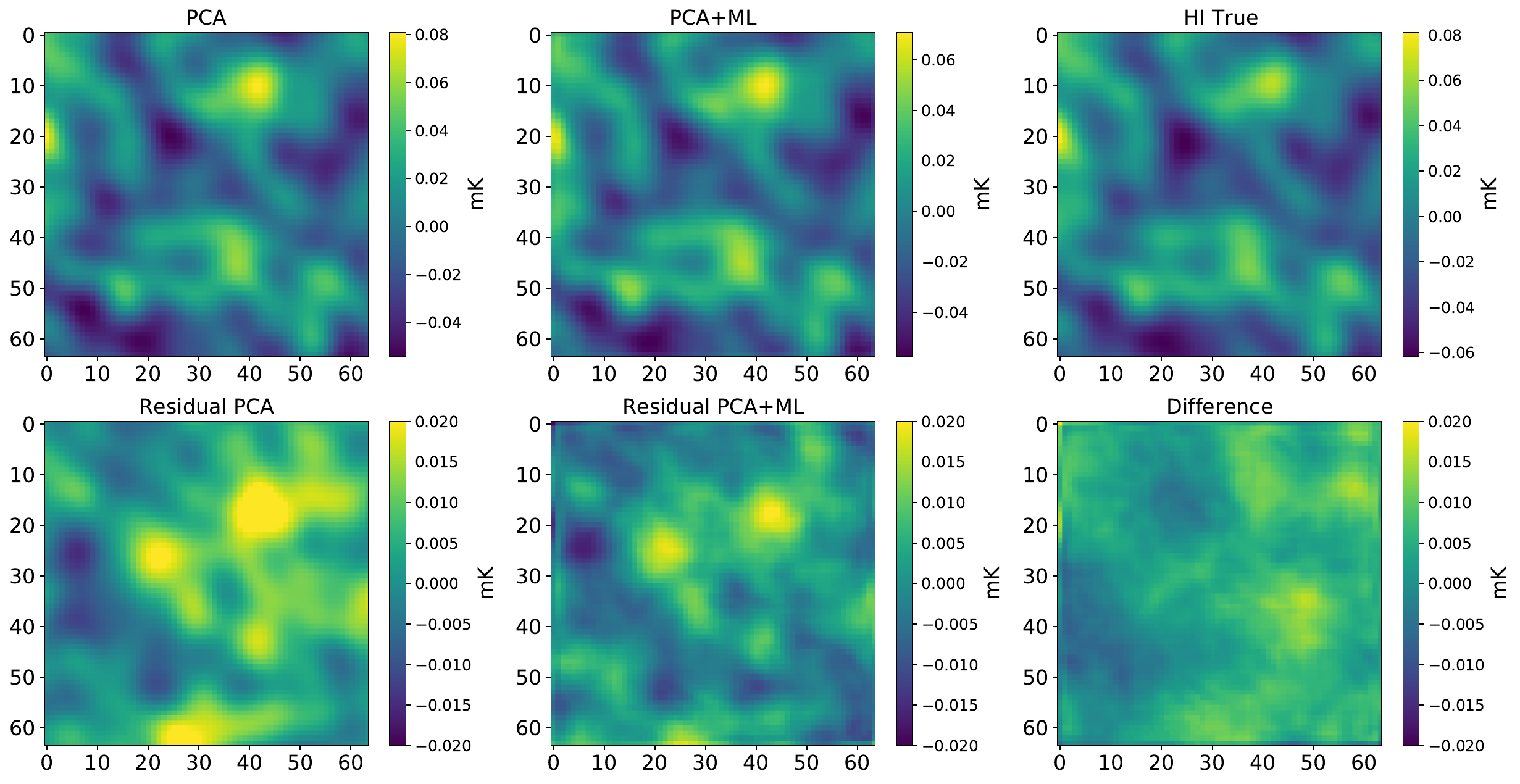}
\caption{The map results of foreground removal under the \textsc{CoLoRe} simulation model.  \emph{Upper:} The reconstructed HI map from PCA  2  alone (\emph{left}), from PCA 2 + ML  (\emph{middle}), and the true (target) HI map (\emph{right}). \emph{Lower:} The residual between reconstructed HI maps from the upper panel and the true HI map. Respectively,  it shows the residual from PCA 2  (\emph{left}), from PCA 2 + ML  (\emph{middle}), and the difference between the two reconstructed maps (i.e., PCA-(PCA + ML); \emph{right}). The plot is an arbitrary sky patch as an example with 64$\times$64 pixels divided from one realisation of \textsc{healpix} map in our test samples. It is from a single frequency channel centred at 862.5\,MHz ($z \approx$0.65).  }
\label{fig:coloremap}
\end{figure*}

\begin{figure}
  \includegraphics[width = 0.97\hsize]{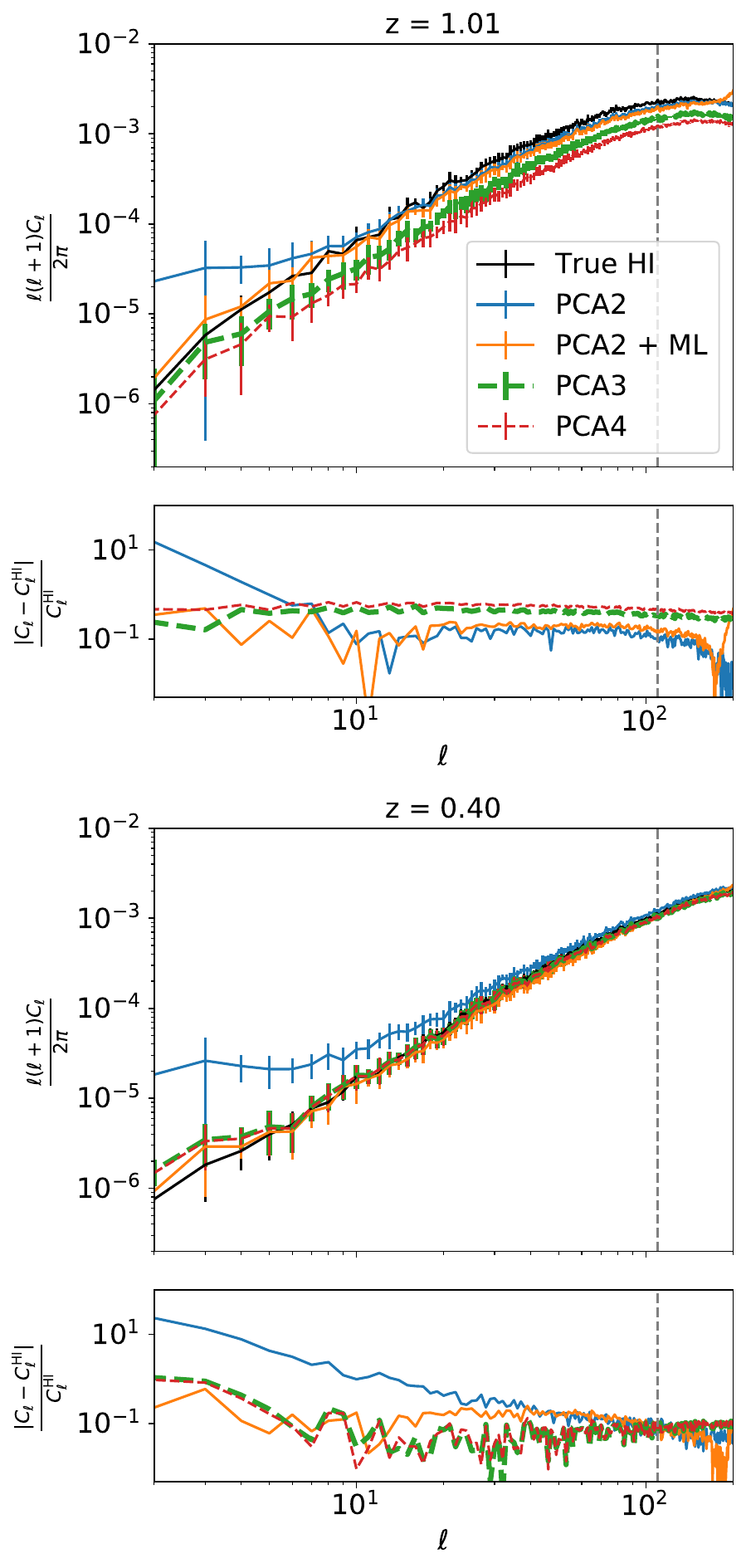}
  \caption{The angular power spectra  of the reconstructed signal and its fractional residual  under the \textsc{CoLoRe} simulation model. The plots are made  the same way as Fig.\,\ref{fig:mscl}.}
\label{fig:colorecl}
\end{figure}
\subsubsection{CoLoRe model}
In this section, we evaluate the U-Net performance under the \textsc{CoLoRe} sky model which has a different HI simulation (see Section\,\ref{sec:color}) but the same foreground simulation with the MS model. We train, validate and test the network consistently with the \textsc{CoLoRe} model.  Fig.\,\ref{fig:coloremap} shows the map results from a $64\times64$ sky patch in the test dataset at a single frequency channel of 862.5\,MHz in this case. The first row shows the output maps from PCA\,2 alone, PCA\,2 + ML and the true HI map respectively. Both output maps  resemble the true HI map in terms of the fluctuating  features and amplitude. The 2nd row shows the corresponding residual map from each method and the difference in the two output maps by subtracting the network output map from the PCA\,2 alone. With the same color bar range, one can see that PCA\,2 has stronger residuals than PCA\,2 + ML. The difference between the two output maps also highlights the large scale diffuse foreground residuals in the PCA\,2  output.

We compare the power spectrum outputs of the \textsc{CoLoRe} model in Fig.\,\ref{fig:colorecl}. The power spectra are computed in the same process as in the MS model. At the higher redshift of $z = 1.01$, the U-Net spectrum closely matches the true HI spectrum, which is confirmed in the lower subpanel with the least fractional residuals from the network. In the lower redshift of $z = 0.40$, the network shows less bias from foreground residuals at large scales ($\ell < 10$) than PCA alone, but is more biased  at small scales ($\ell > 10$) due to signal loss. Nevertheless, the network performance is  comparable with PCA\,3 and PCA\,4 alone in the lower redshift. In both redshifts,  the network has an average fractional residual of $\sim10\%$ of the signal over all scales. This is consistent with the results from the MS model.

The $R^2$ scores in the \textsc{CoLoRe} model are computed following the same procedure as in the MS model, and are shown in the middle panel of Fig.\,\ref{fig:modelr2}. Similar to the MS model, the PCA\,2 + ML has the highest $R^2$ scores over all frequencies. This can be seen both  in comparing the $R^2$ score curves the upper subpanel and  the positive  $R^2$ score difference in the lower subpanel. We again observe the drop of $R^2$ score values at the two ends of frequency channels due to the lack of continuous frequency information for PCA.

Overall, the network results under the \textsc{CoLoRe} model are consistent with those from the MS model. The visual inspection of the sky maps both show visible subtraction of foregrounds further from the PCA\,2 pre-processing. The fractional residual in the power spectra are constantly around $10\%$ of the signal. The $R^2$ scores of the network surpass those from PCA alone in both cases thanks to the 3D convolution operations. These results demonstrate that as long as the training and testing data samples are consistently from the same sky model, the network is not sensitive to  the change of HI signal models and has consistent performance over different signal models. 

\begin{figure*}
\includegraphics[width = 0.97\hsize]{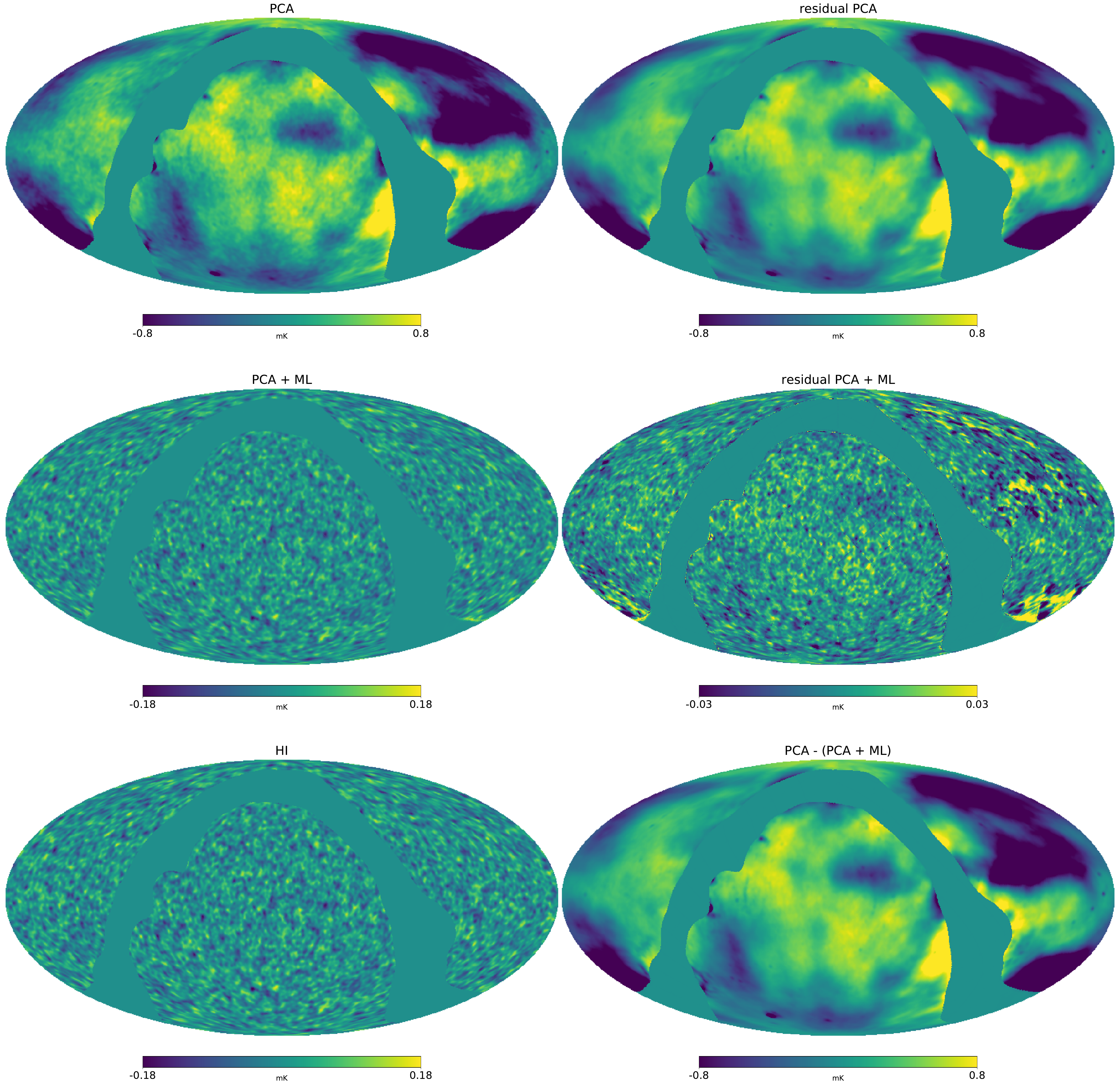}
\caption{The map results of foreground removal under the Planck sky (PSM) simulation model. Full sky \textsc{healpix} maps are shown here to demonstrate the ability of the network to handle mask. From top to bottom and from left to right, the panels are respectively the reconstructed HI map from PCA\,2 alone,  the residual map from PCA\,2 mode alone,  the reconstructed HI map from PCA\,2 + ML,  the residual map from PCA\,2 + ML,  the original true HI map, and   the difference between the two reconstructed maps from PCA\,2 alone and PCA\,2 + ML. The plots are from a single realisation of \textsc{healpix} maps in the test dataset, centred at a single frequency channel at 862.5\,MHz ($z \approx$0.65).  }
\label{fig:psmmap}
\end{figure*}

\begin{figure}
  \includegraphics[width = 0.97\hsize]{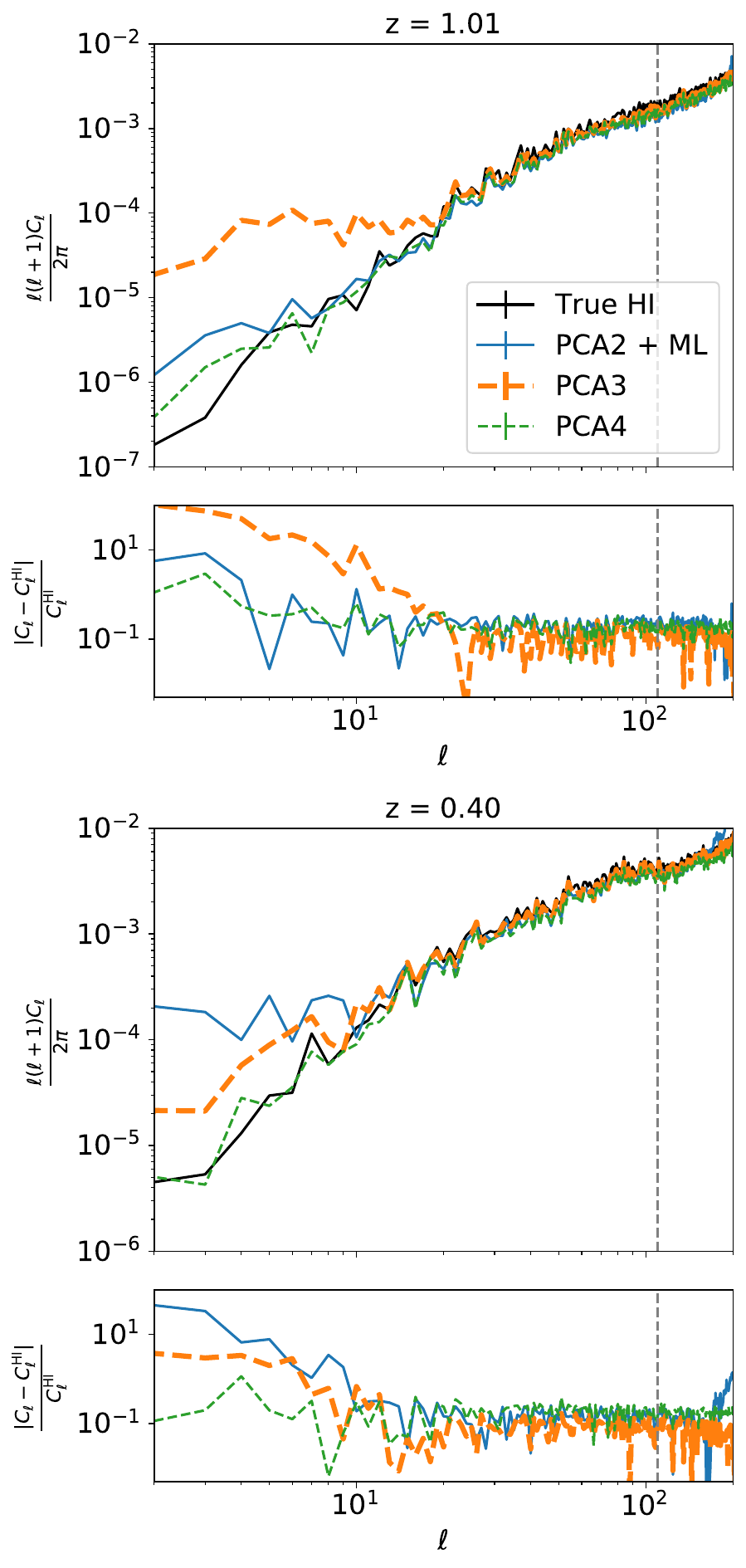}
\caption{The angular power spectra of the reconstructed signal and its fractional residual under the PSM  simulation model. The plots are made the same way as Fig.\,\ref{fig:mscl}. PCA 2 alone is not shown here as it alone does not reduce the foreground enough to recover the signal reasonably close to the true signal. In each case, the power spectrum is computed from the single  \textsc{healpix}  test sample  as the PSM model provides only one realisation of the foregrounds. }
\label{fig:psmcl}
\end{figure}
\subsubsection{PSM model}
To understand whether the U-Net is sensitive to the foreground model, we train, evaluate and test the network consistently under the PSM sky model. The PSM model replaces the foreground components in the MS model by observed templates but keeps the same HI simulation.  Note that in the PSM model, our test dataset  contains only one realisation of \textsc{healpix} map (equivalently 192 test samples) based  on the real  observed templates (see Section\,\ref{sec:psm}).  Fig.\,\ref{fig:psmmap} shows the outputs reassembled to the full \textsc{healpix} sky map at the frequency channel of 862.5\,MHz.   We apply the 20$\%$ \emph{Planck} Galactic mask to the  data since the PCA\,2 alone pre-processing does not reduce the dynamic range enough in the Galactic plane region for the network.  The Galactic mask is applied consistently to training, validating and testing  \textsc{healpix} maps after applying PCA\,2 pre-processing and before dividing into $64\times64\times64$ sky cubes. The masked region is shown in uniformly green  in Fig.\,\ref{fig:psmmap}. The left column shows the maps from  PCA\,2, PCA\,2 + ML and the true HI map. The right column shows the corresponding residuals and the difference between the two output maps. It is reassuring to see that the network can handle masked data without additional prior information beyond training.

As can be seen in the middle left panel of Fig.\,\ref{fig:psmmap}, the network correctly reconstructs  the mask after its denoising procedure with only limited confusion at the boundary of the masked region.  From the first row of Fig.\,\ref{fig:psmmap}, the PCA\,2 pre-process has significant foreground residuals shown as bright diffuse features with strong emission as can be seen from the color bar range. In contrast, the network output map (\emph{middle left}) successfully recovers the HI fluctuation features to the same amplitude level as the true map. The still bright features in the residual map from PCA\,2 + ML (\emph{middle right}) yet indicate foreground residuals, however,  at a much smaller amplitude compared to the signal. The difference between the two output maps (\emph{bottom right}) confirms the dominating foreground contamination left by the PCA\,2 pre-process, which has been largely removed by the U-Net. 

Fig.\,\ref{fig:psmcl} shows the power spectra of the true HI and reconstructed maps from the PSM model. The power spectra are computed from the reassembled single test \textsc{healpix} map with the mask effect corrected. We do not show the  PCA\,2 alone pre-process results because the spectra are far-off from the true spectrum as expected from the map in Fig.\,\ref{fig:psmmap}. At the higher redshift bin of $z = 1.01$, the spectrum from the U-Net is comparable with that of PCA\,4. Compared with PCA\,3, PCA\,2 + ML has significantly reduced large-scale bias due to foreground residuals. Based on the fractional residual in the lower subpanel, the residual from the network is comparable with the signal at large scales of $\ell<5$ and drops back to $\sim10\%$ of the signal at scales of $\ell\gtrsim10$.   In the lower redshift bin, PCA\,4 gives the optimal results with the mostly matching spectrum towards the truth and the least fractional residual. PCA\,2 + ML, however, is less competitive with large  bias due to foreground residuals left by the PCA\,2 pre-processing at large scales of $\ell\lesssim10$.  Nevertheless, the fractional residual of PCA\,2 + ML drops to $\sim10\%$ of the signal at $\ell\gtrsim10$.  Based on both redshift cases,  the network results are indeed less sensitive to the redshift as in the other two sky models. Therefore, while PCA\,3 and PCA\,4 benefit from the reduced foreground to signal ratio at low redshift, the U-Net does not have such relevant knowledge and thus loses its advantage  to PCA alone at low redshift. In this case, the U-Net provides an intermediate solution between PCA\,3 and PCA\,4, with better reconstruction at higher redshift than PCA\,3 and worse reconstruction at lower redshift than PCA\,4. 

Compared with the MS and \textsc{CoLoRe} models,  the larger bias from the network  under the PSM model is due to the less effective pre-process with PCA\,2. The input data into the network in the PSM model has stronger foreground residuals, making it more difficult for the network to reconstruct the signal.  However, we warn that the power spectra  in the PSM model are computed from a single simulation of \textsc{healpix} test map.  One thus needs to be aware of the potential large sample variance in the spectra during interpretation, in particular, at large scales with possible impacts from the mask.  We also tested the network with PCA\,3 pre-processing and found that  the network does  not show much improvement beyond PCA\,3. This is because the PCA\,3 pre-process already significantly reduce the foregrounds such that the input data and the target signal for the network is too similar to be easily  distinguished and thus results in ineffective training for denoising.

The $R^2$ scores of the test \textsc{healpix} map from different methods are shown in the right panel of Fig.\,\ref{fig:modelr2}. The PCA\,2 pre-processing does not give  sensible values, as expected from the map results in Fig.\,\ref{fig:psmmap}. The $R^2$ score from PCA\,2 + ML  highlights the challenge to reconstruct the HI map from the PCA\,2 pre-processing to a reasonable level comparable to PCA\,3 and PCA\,4. However, the network shows lower  $R^2$ score than PCA alone with 3 or 4 mode removal as demonstrated by the negative $R^2$ score difference in the lower subpanel. This is attributed to the reduced efficiency from the pre-process step as already reflected in the power spectrum result. 

Based on the results from all the three sky models above, the network performance does not vary significantly among different models. In all cases, the network manages to further subtract foregrounds from the PCA\,2 pre-processed data. The yielded power spectra and $R^2$ scores from PCA\,2 + ML are comparable with PCA\,3 or PCA\,4 alone.  Compared with PCA alone, the network results are always insensitive to the redshift due to its lack of knowledge about the change of the foreground to signal ratio along redshift. The 3D convolution operations in the U-Net ensure  the accurate signal reconstruction in the map (image) space.  However, the network relies on the pre-processing step and can be limited by the pre-processing results.  Overall, the performance of our U-Net is stable under different sky models so long as the network is retrained for each sky model. It provides complementary results to PCA alone with certain advantages depending on the particular dataset at particular redshift.  

\begin{figure}
  \includegraphics[width = 0.97\hsize]{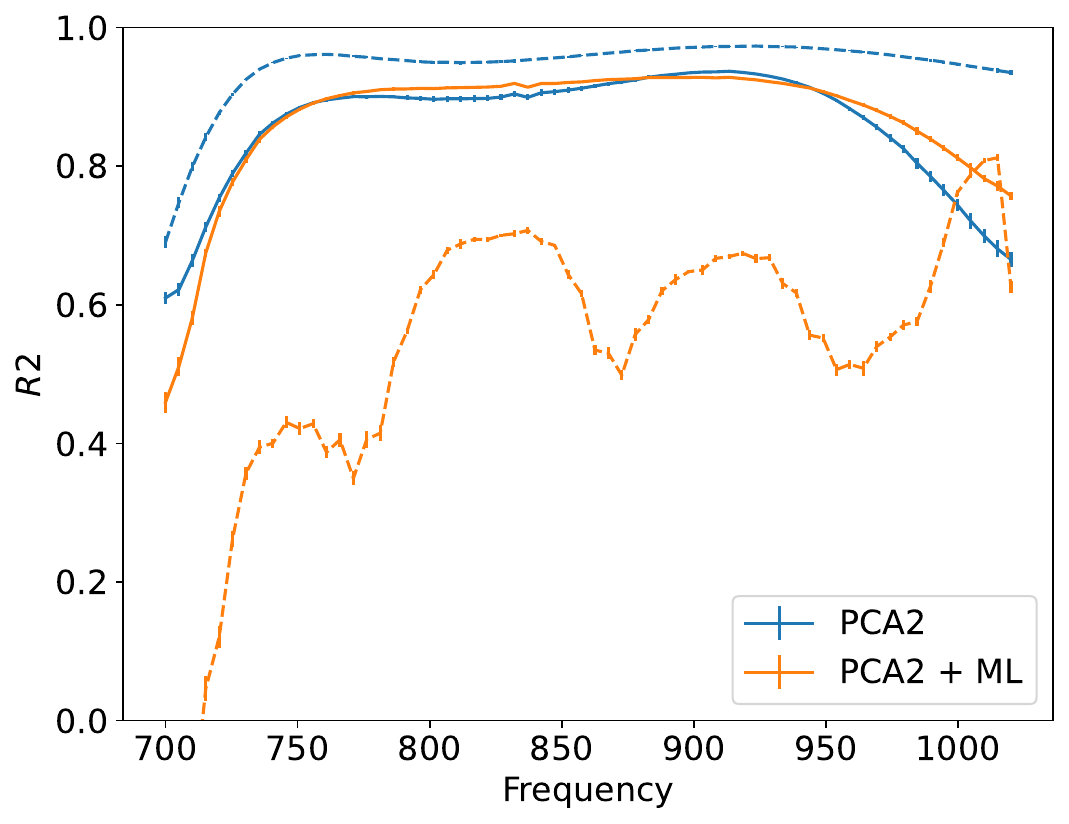}
\caption{The $R^2$ scores of the reconstructed signals when the training and testing dataset are from different sky models. The \emph{solid } lines have the network trained by the MS model but tested by the CoLoRe model. Vice versa, the \emph{dashed} lines have the network trained by the CoLoRe model but tested by the MS model. In each case, we compare the results from PCA\,2 + ML with PCA\,2 alone.}
\label{fig:clcross}
\end{figure}
\subsubsection{Cross-model training}
During our test of the U-Net performance under the above three sky models, we always  train, validate and test the network consistently with the same sky model but different realisations. In addition, we also investigate whether the network can be trained on one model but successfully denoise data from a different model, which we call the ``cross-train'' test. This is a good test of the {\em generalizability} of a deep neural network, its ability to adapt to data never seen during training. We have tried all possible combinations of training and testing among the three sky models.  None of our cross-trained U-Net models are able to outperform PCA.  As an example, in Fig.\,\ref{fig:clcross}, we show the $R^2$ scores  of PCA\,2 alone and PCA\,2 + ML when the network is trained with the MS model but tested with the \textsc{CoLoRe} simulated data (\emph{solid}) and vice versa (\emph{dashed}).

In the \emph{solid} lines in  Fig.\,\ref{fig:clcross}, PCA\,2 + ML has very similar $R^2$ scores to PCA\,2 alone. Compared with the middle panel of Fig.\,\ref{fig:modelr2} where the network is trained and tested consistently by the \textsc{CoLoRe} model, the $R^2$ scores of PCA\,2 + ML  in this cross-model case are  much worse. In the \emph{dashed} lines in Fig.\,\ref{fig:clcross}, the $R^2$ scores from PCA\,2 + ML are incomparable to PCA\,2 alone and significantly worse than the left panel of Fig.\,\ref{fig:modelr2}, where the network is consistently trained and tested under the MS model. In both case of Fig.\,\ref{fig:clcross}, there is no evident  benefit to use the network since its performance is no better than simply PCA alone. We do not show further  results from other cross-model combinations here since they are very similar to Fig.\,\ref{fig:clcross}. During our cross-model tests, we do not statistically observe a clear tendency on which types of model combination still allows decent signal reconstruction from the U-Net. The performance of the network  is always worse when cross-trained rather  than being trained and evaluated consistently under the same model. 

The degraded network performance in the cross-model case is somewhat expected since the three sky models are very different from each other. The network cannot simply denoise a dataset without relevant prior training or knowledge about it.  Therefore, in order to achieve the optimal results using a U-Net, one should have  reasonable knowledge about the data and thus train the network accordingly with appropriate training samples. We thus warn that one should be cautious when applying the DL network to real 21cm data, as the simulation used for network training is likely to under-represent the complexity of the real Universe.

\subsection{Frequency structure dependency}\label{res:sys}
In this section, we study the robustness of the U-Net against  systematics-induced frequency distortions. We test our network on two of the most common frequency-dependent instrumental effects: i) a frequency-dependent beam; ii) a frequency dependent gain drift. In each case, we first input the systematics contaminated test data directly into the above trained network  which has no knowledge about the  systematics. Secondly, we re-train the network with systematics corrupted training samples and see if the denoising results in this case are better than without re-training.  We adopt the MS sky model throughout this section, since we have seen that the network is stable under different sky simulations.

\begin{figure}
  \includegraphics[width = 0.97\hsize]{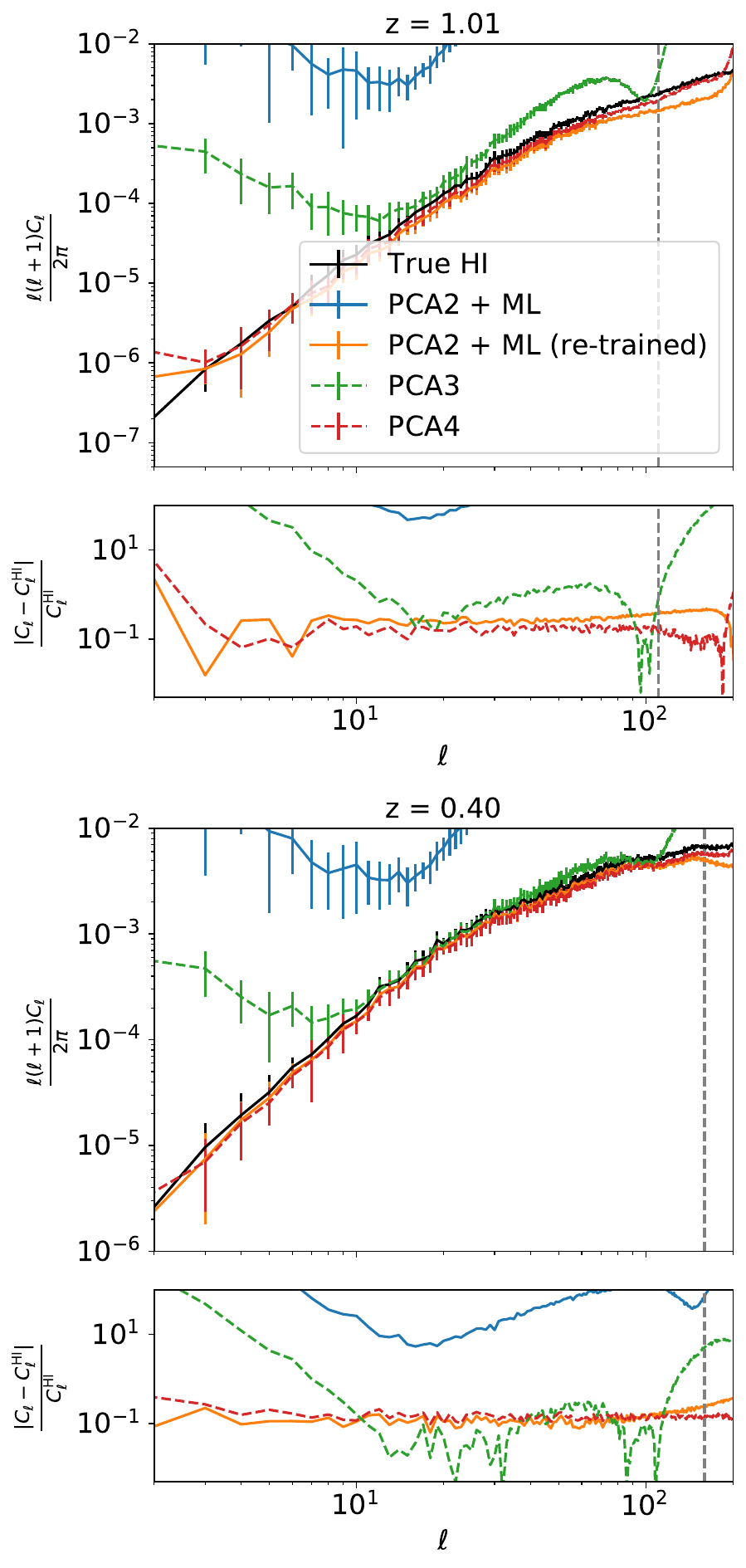}
   \caption{The angular power spectra  of the reconstructed signal and its fractional residual in the case of frequency-dependent beam data. The plots are made the same way as Fig.\,\ref{fig:mscl}. Two network reconstructed results are shown in this case. \emph{Blue}: The network is the same as in the MS model  without prior training of the frequency-dependent beam. \emph{Orange}: The network is re-trained with frequency-dependent  beam  data. We compare the network results with PCA 3 and 4 mode removal. }
\label{fig:msbeamcl}
\end{figure}

\begin{figure}
  \includegraphics[width = 0.97\hsize]{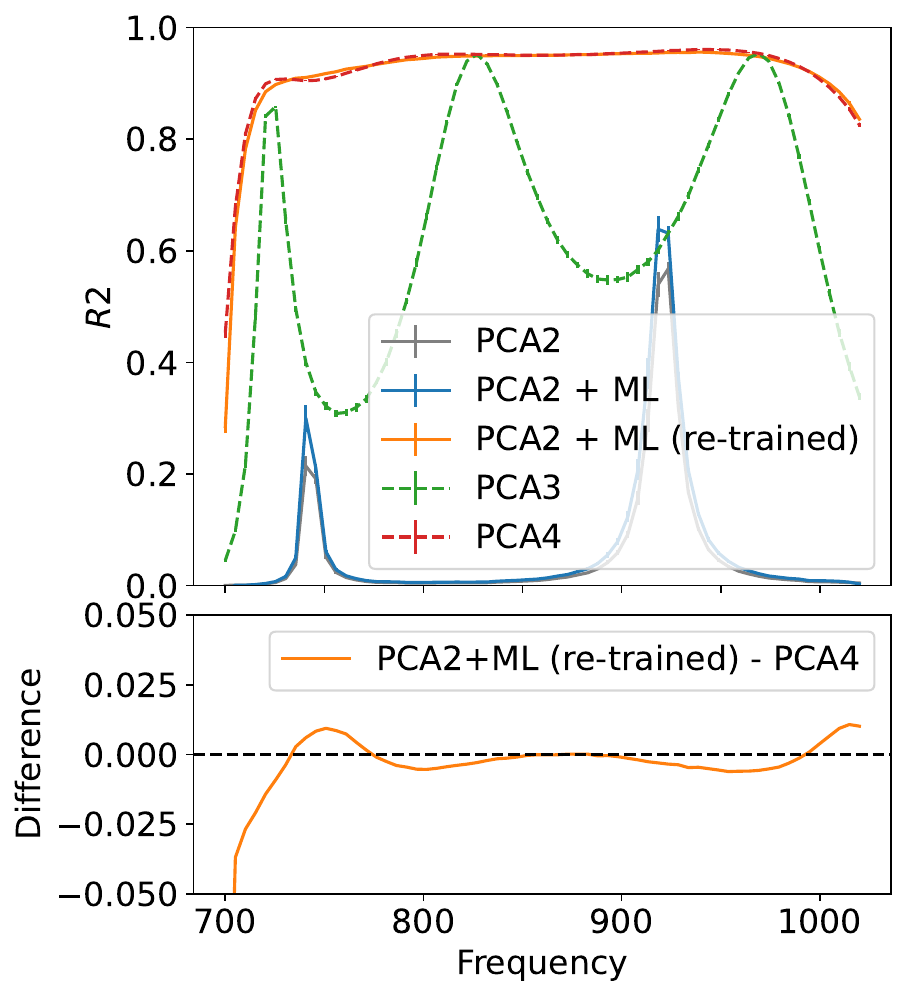}
  \caption{The $R^2$ scores of the reconstructed signal in the case of  beam-varying data. In the upper panel, we show the $R^2$  scores from two network results: with and without re-training using  the frequency-dependent beam data. In comparison, we show the results from PCA 2, 3, and 4 mode removal. In the lower panel, we show the difference between the two best $R^2$ scores from retrained PCA 2 + ML and PCA 4, in order to evaluate the performance of retrained network in comparison with PCA alone. }
\label{fig:beamr2}
\end{figure}

\subsubsection{Frequency-dependent beam}\label{secbeam}
We  apply the frequency-dependent beam from Equ.\,\ref{equ:beam} into the test dataset from the MS model. We note that the beam shape is still assumed to be Gaussian except its FWHM evolves along the frequency direction. We did not consider more complex beam shapes because \cite{nlg+22} already tested a similar U-Net  with a Cosine beam model and found that the network reconstructs the signal better than PCA alone in such cases. Instead, our work focuses on the impact of beam-induced frequency structures on the network and accordingly, the correct approach to employ the DL network in such cases.

At first, we directly denoise the test data convolved with frequency-dependent beams by the network keeping the weights fixed from the initial training  in Section\,\ref{sec:ms}. The network is trained by data from the MS sky model with fixed beam resolution at the highest frequency channel. Therefore, the network in this case has no prior information about the beam-induced frequency structure in the test dataset. The test data are also pre-processed with PCA\,2 following the same procedure as in the MS model.  We show the power spectrum results of DL along with PCA\,3 and PCA\,4 alone in Fig.\,\ref{fig:msbeamcl}. In both redshift bins, we can see that the power spectra from PCA\,2 + ML (\emph{blue})  are far off the true spectrum. PCA\,3 (\emph{green}) alone gives better results than PCA\,2 + ML but its reconstructed power spectra are still significantly biased. The spectra from PCA\,4 (\emph{red}) best match the true spectrum. Compared with the fixed-beam in Fig.\,\ref{fig:mscl}, the frequency-dependent beam completely distorted the performance of the network.

In the second case, we also convolve the training and validation dataset from the MS model with frequency-dependent beams. The network is re-trained from scratch using these updated data. The target in this case is the HI cubes with evolving resolutions along the frequency direction.  We apply the re-trained network on the test data with frequency-dependent beams. This is to understand if by re-training with the beamed data, the network can successfully recognize the beam-induced frequency structure and improve its signal reconstruction in this case. The power spectrum of the re-trained result is shown in \emph{orange} in Fig.\,\ref{fig:msbeamcl}. In both redshift bins, the re-trained power spectra closely match the true spectra. In the lower subpanels, the re-trained network reduce the fractional residuals back to $10\%$ of the signal, consistent with the case in Fig.\,\ref{fig:mscl} with fixed beam. These results demonstrate that re-training the network with the appropriate beam  is crucial to correctly denoise data with frequency-dependent beams.

We also compare the results in the map space by plotting the $R^2$ scores from different methods in Fig.\,\ref{fig:beamr2}. In the upper panel, we can see that without re-training the network,  PCA\,2 + ML (\emph{blue})  barely improves the $R^2$ score compared with the pre-processing using simply PCA\,2 alone. Similar to the power spectrum results in Fig.\,\ref{fig:msbeamcl}, PCA\,3 alone also struggles to accurately reconstruct maps with sensible $R^2$ scores. The re-trained network (\emph{orange}), however, significantly improves the signal reconstruction results with comparable $R^2$ scores as PCA\,4 alone.  In the lower panel, we plot the difference in  $R^2$ score between the retrained  network and PCA\,4  for comparison. Indeed, the two methods return $R^2$ scores with minimal difference, indicating that their output maps have similar accuracies. Comparing the results from the network with and without re-training, the $R^2$ scores once again demonstrate the importance of training the network with the prior relevant  information to proper denoise frequency-dependent beamed data.

From both the power spectrum and the $R^2$ score analysis, we have seen that the U-Net cannot handle data corrupted by the frequency-dependent beam without learning the relevant information in advance. However, once the network is trained consistently with the same beam, it can reconstruct the data  as accurately as if there is no such beam effect.  The additional prior beam information during training is crucial to maintain a decent network performance.

\begin{figure}
  \includegraphics[width = 0.97\hsize]{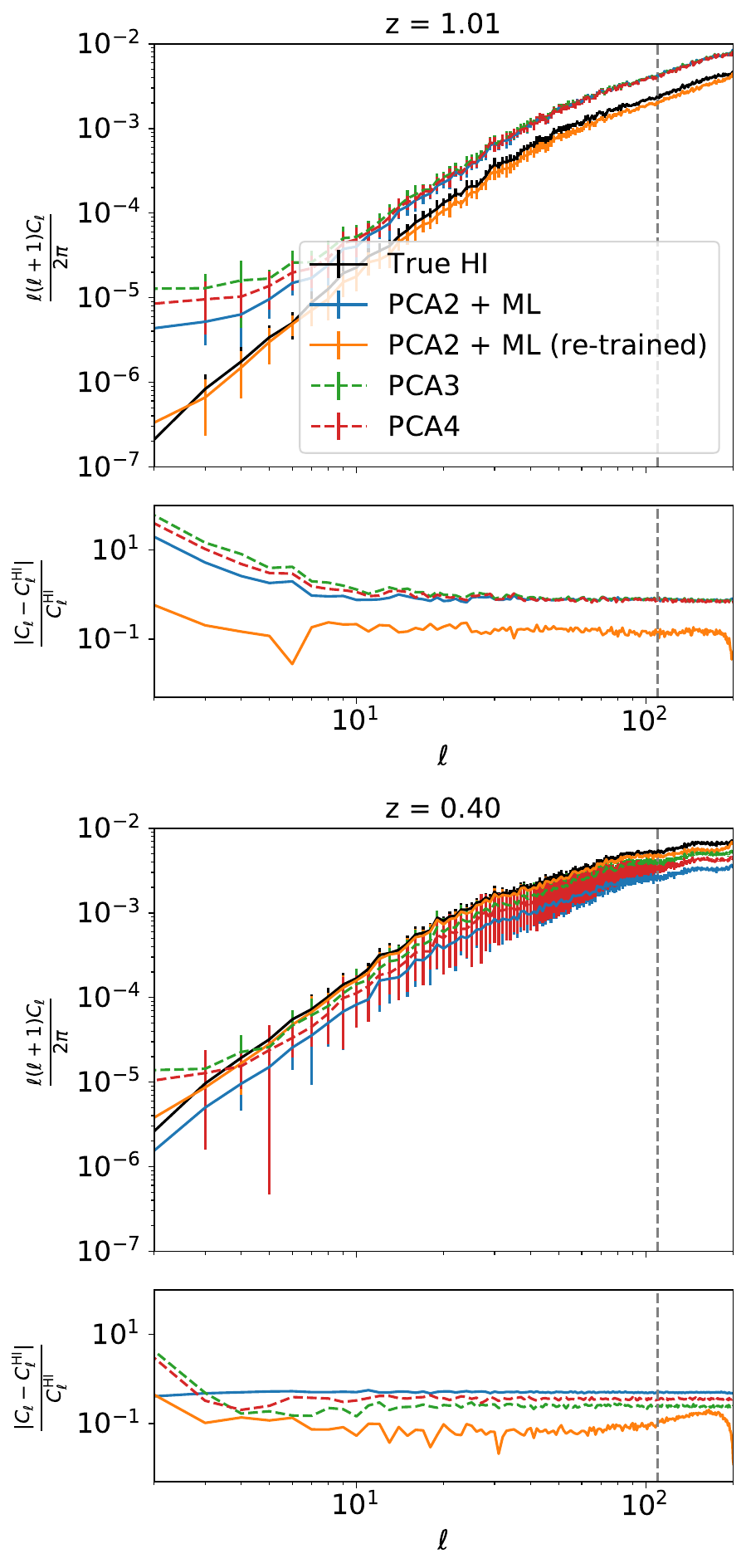}
  \caption{The angular power spectra  of the reconstructed signal and its fractional residual in the case of frequency-dependent gain drift. The plots are made the same way as Fig.\,\ref{fig:msbeamcl}. Two network results are shown in this case. \emph{Blue}: The  network is the same as in the MS model  without prior training of the frequency-dependent gain drift. \emph{Orange}: The network is re-trained with data suffering from frequency-dependent  gain drift. We compare the network reconstructed results with PCA 3 and 4 mode removal.}
\label{fig:MSsingaincl}
\end{figure}

\begin{figure}
 
  \includegraphics[width = 0.97\hsize]{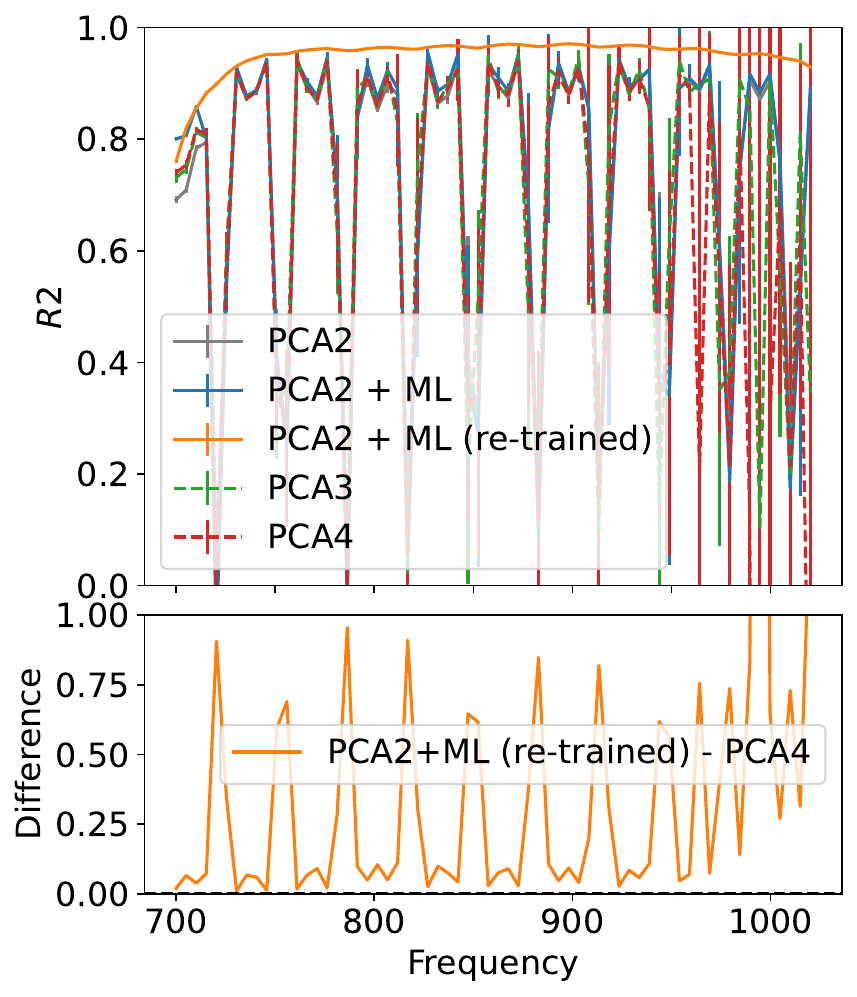}
  \caption{The $R^2$ scores of the reconstructed signal in the case of  frequency-dependent gain drift. The plots are the same as Fig.\,\ref{fig:beamr2}. In the upper panel, we show the $R^2$  scores from two network results: with and without re-training using  the frequency-dependent beam data. In comparison, we show the results from PCA 2, 3, and 4 mode removal. In the lower panel, we show the difference between the two best $R^2$ scores from retrained PCA 2 + ML and PCA 4.}
\label{fig:singainr2}
\end{figure}
\subsubsection{Frequency-dependent gain drift}\label{sec:res-singain}
Following the above procedure in section\,\ref{secbeam}, we study the network performance at the presence of frequency-dependent gain drifts. Gain drifts are applied to data cubes before PCA\,2 pre-processing following Equ.\,\ref{equ:gain}\,\&\,\ref{equ:gv}. We adopt the MS sky model with fixed beam in this case to avoid other confounding factors.  We firstly only apply  gain drifts in the test dataset and  input into the network without re-training.  The network thus has no knowledge about the gain drifts in the test dataset.  In the second case, the network is re-trained by introducing  gain drifts into the training and validation samples.  This is to investigate whether introducing the gain drift information  during re-training will improve the network performance. When applying gain drifts, as described in section\,\ref{sec:gain}, we vary the parameters of the gain drifts with a Gaussian distribution so that each realisation has a slightly different gain drift.

The power spectrum results are shown in Fig.\,\ref{fig:MSsingaincl}. One can see that at the higher redshift of $z = 1.01$ (\emph{upper}), the spectrum from  PCA\,2 + ML without re-training (\emph{blue}) shows excess of power compared with the true spectrum (\emph{black}), indicating residual  foregrounds in the data. At the lower redshift of $z = 0.4$ (\emph{lower}), PCA\,2 + ML without retraining  results in over-subtraction as its power spectrum is lower than that of the true signal. Similarly, PCA\,3 and PCA\,4 alone also show excessive of power at higher redshift and under power at low redshift. In contrast, after re-training the network with the gain drift information, its power spectra (\emph{orange}) closely match the true spectra in both redshifts. From the lower subpanels, the re-trained network returns a fractional residual of $\sim10\%$ of the signal over all scales at both redshifts. This is consistent with the original case of the MS model in Fig.\,\ref{fig:mscl} without gain drifts.

The $R^2$ scores of different methods are shown in Fig.\,\ref{fig:singainr2}. The obvious feature is that except the re-trained network (\emph{green}), all other methods have  strongly oscillating $R^2$ scores along the frequency.   This is due to the sinusoidal oscillation of the gain drifts as in Equ.\,\ref{fig:singainr2}. A similar  case has been reported in \cite{mss+21}, where they observe oscillating patterns in the resultant 21cm power spectrum when they introduce a frequency-dependent cosine beam model in their simulated data. In contrast, the retrained network still gives stable $R^2$ scores along the frequency with a sensible value of $R^2\sim0.9$. This highlighted in the lower panel where as an example, the difference between the re-trained network  and PCA\,4 alone shows sinusoidal-like features. 

Once again, our results show that re-training the network with the gain drift information is crucial to properly reconstruct the signal. The re-trained network produces more accurate results than PCA alone in both the map space and the power spectrum space. Without the re-training, the network is not able to correctly handle data with ``unexpected'' frequency-dependent gain drifts.    Our tests of both the frequency-dependent beam and the gain drift  warn us that before applying a DL network, one should have some reasonable knowledge about the data, such as the type of systematics residing in it. The network must be trained accordingly for a reasonable performance. Simply training the network on one set of simulation and expecting it to denoise data with arbitrary systematics will not work.  On the other side, we also apply the re-trained network with systematics information to the original test data without systematics. This is to understand if one can simply train the network with all possible systematics models and then denoise any data. We found that in this way, the network results in over-subtraction due to the network over-fitting. We thus address the importance of knowing the data before hand and appropriately training the network accordingly.

\section{Conclusions}\label{dissec}
In this work, we test the robustness of DL for 21cm foreground removal against different simulation models and instrumental systematics. We use a common U-Net as our network for removing foregrounds. We train and test the network under three different sky models based on pure Gaussian realisations (baseline model), the Lagrangian perturbation theory and observed templates of the real sky respectively. In each case, we evaluate the network performance through map visualisation, power spectrum comparison and $R^2$ score accuracy. In the second  part of the paper,  we introduce a frequency-dependent Gaussian beam and a sinusoidal gain drift respectively into our data to investigate the impact of systematics on the network performance. The data in this case are simulated using the baseline Gaussian model. We compare the results from  networks trained with and without systematics-corrupted data to study whether introducing systematics information during training is important for the network performance. In all cases, the data are pre-processed with PCA\,2 mode removal before inputting into the network to reduce the large dynamic range between foregrounds and the signal, which the network struggles to handle. 

In the sky model test, we train and test the network consistently under each model and the network gives consistent results across the three sky models. On average, the residual foreground from the network  is   $\sim10\%$ of the signal over all angular scale at our considered frequency range of SKA-MID Band\,1based on the power spectrum analysis. While continuing efforts are being made to match simulations to real observations,  it is useful to learn that U-Net already performs consistently across existing  simulations before applied to upcoming 21cm data. By testing the network with the Planck sky model, we found that the network can be applied to masked data. As long as the network has been trained with masked data, it can predict masked regions  without introducing unwanted features. We also tried to train the network with one sky model but test on another model to see if the cross-trained network can denoise data from a different model. Our results show that the network fails to correctly reconstruct the signal in such cases. This means that the training and testing data has to be consistent for a decent performance of the network. Therefore,  with this type of DL network, one needs almost perfect knowledge of the properties in the data to use the network efficiently. However, in reality, it is very difficult to know the data precisely as we will be obtaining 21cm data almost for the first time. We suggest comparing reconstructed signals from different methods  to properly interpret the results in the case of real data. 

As a relatively new technique for foreground removal,  DL network also gives comparable results with traditional approaches such as PCA in the case of no systematics. In comparison, PCA is more sensitive to the evolution of foreground to signal ratio along the redshift,  while our network is not trained with such information and thus its results are independent of the  redshift. In addition, the 3D convolution operations in the U-Net  are designed to match the output cube into the true HI cube. Therefore, the output map from the network has better accuracy than that from  PCA in the image space as demonstrated in our $R^2$ score analysis.  Nevertheless, the network results agree with PCA results with small differences at certain angular scales and redshifts depending on the specific cases. DL  network thus has the potential  as a complementary method to traditional approaches for 21cm foreground removal. This will be useful once real data are available where one can compare the results from a DL network and PCA to obtain the best signal reconstruction, especially since it is unclear to  decide the optimal number of mode removal for PCA in the case of real data.

At the presence of systematics, the network fails to reconstruct the signal if it is not trained consistently with the systematics-corrupted data. In other words, the network in our framework is not able to denoise data with ``unexpected'' systematics. However, once we introduce the same type of systematics into the training dataset and re-train the network,  its performance is significantly improved back to the scenario of no systematics. The residual foregrounds in the network reconstructed map  drop back to $\sim10$ of the signal. This is the case for both the frequency-dependent beam and gain drifts. These results demonstrate that the prior information on the systematics during training is critical for the network to successfully  reconstruct systematics contaminated data.  The network requires to learn at least the general structure of the certain types of systematics to correctly perform signal reconstruction. For example, in our case the network correctly reconstructs the signal from data with gain drifts even its training samples have different gain drift values. In another scenario,  we apply the systematics-trained network to denoise data without systematics in order to understand if one can simply train the network with all possible systematics and blindly apply it to the data. However, we found that the network will result in over-subtraction in such case due to the network over-fitting. We thus address the importance of having a decent knowledge of the data and consistently train the network accordingly.   In the case of real observations, certain information of systematics can potentially  be obtained in advance such as the instrumental  beam response  and, to some degree, the  gain fluctuations.  In such cases, we suggest a first inspection of the obtained data to estimate the potential systematics, and use those prior information during network training to obtain the optimal network performance. Nevertheless, cautions must be taken when directly applying the network to real 21cm data since the simulated training samples are unlikely to  fully represent the complexity in the upcoming  21cm data. 

In summary, U-Net is  a stable new technique for 21cm foreground removal, complementary to traditional approaches. However, like all other methods, it has certain limitations and one should take some actions  when applying the network for foreground removal for the optimal results. For example, its performance heavily relies on the prior information about the data and thus one should provide such priors either through the training samples or directly into the network structure. It also relies on a pre-processing step to reduce the large dynamic range between foreground and signal, which could result in already over-subtraction before the U-Net step. One possible way to overcome this limitation might be to try the diffusion network in the future, which gradually reduce the noise level over several steps \cite{kee+22}.  In addition, our network does not make use of the physical information of the foreground to signal ratio at different redshifts. One potential improvement in the future would be to incorporate such information in the network structure to further help the network with the denoising. For cosmology analysis that relies on accurate 21cm power spectrum reconstruction, perhaps a future improvement would be to implement power spectrum calculation in the network structure and return double outputs to match not only the target cube but also the target power spectrum.

\section*{Acknowledgements}
TC, MB and MS acknowledge the financial support from the SNSF under the Sinergia Astrosignals grant (CRSII5\_193826).  This work was supported by EPFL through the use of the facilities of its Scientific IT and Application Support Center (SCITAS).

\section*{Data availability}
The dataset under the MS sky model can be simulated using the github repository: \url{https://github.com/SharperJBCA/SWGSimulator}. The \textsc{CoLoRe} simulation package is available at \url{https://github.com/damonge/CoLoRe}. The code for the Planck sky model is accessible at \url{https://bitbucket.org/IMpipeline/pipeline}. The U-Net used in this paper is available upon request to the author.

\appendix
\section{U-net architecture}\label{app:unet}
Our network has a commonly used U-Net structure \citep[e.g.,][]{rfb15, mlv+21, nlg+22, gln+22}. Table\,\ref{tab:architect} gives the detailed architecture of the network with different layers. 
\begin{table}
  \begin{tabular}{l|l|l}
    \hline
    \hline
    Layer ID & Layer type & Output dimension \\
\hline
1& InputLayer  & (64,64,64,1) \\
2 & 3$\times$Conv3D\_Blocks  & (64,64,64,32)    \\
3        & MaxPooling3D & (32,32,32,64)    \\
4        & Dropout & (32,32,32,64)    \\
5        & 3$\times$Conv3D\_Blocks & (32,32,32,64)    \\
6        & MaxPooling3D & (16,16,16,128)   \\
7        & Dropout & (16,16,16,128)   \\
8        & 3$\times$Conv3D\_Blocks & (16,16,16,128)   \\
9        & MaxPooling3D & (8,8,8,256)      \\
10       & Dropout  & (8,8,8,256)      \\
11       & 3$\times$Conv3D\_Blocks & (8,8,8,256)      \\
12       & MaxPooling3D & (4,4,4,512)      \\
13       & Dropout  & (4,4,4,512)      \\
14       & 3$\times$Conv3D\_Blocks  & (4,4,4,512)      \\
15       & 3$\times$Conv3DTrans\_Blocks & (8,8,8,256)      \\
16       & Concatenate & (8,8,8,512)      \\
 &  {[}Layer 11, Layer 15{]} & \\
17       & Dropout  & (8,8,8,512)      \\
18       & 3$\times$Conv3DTrans\_Blocks & (16,16,16,128)   \\
19       & Concatenate & (16,16,16,256)   \\
 &  {[}Layer 8, Layer 18{]} & \\
20       & Dropout & (16,16,16,256)   \\
21       & 3$\times$Conv3DTrans\_Blocks   & (32,32,32,64)    \\
22       & Concatenate  & (32,32,32,128)   \\
 &  {[}Layer 5, Layer 21{]} & \\
23       & Dropout  & (32,32,32,128)   \\
24       & 3$\times$Conv3DTrans\_Blocks  & (64,64,64,32)    \\
25       & Concatenate  & (64,64,64,64)    \\
 &  {[}Layer 2, Layer 24{]} & \\
26       & Dropout  & (64,64,64,64)    \\
27       & Conv3D  & (64,64,64,1)     \\
\hline
  \end{tabular}
 \begin{tabular}{c|c@{\hspace{43.5pt}}|c}
   Conv3D\_Blocks & &Conv3DTrans\_Blocks \\
   \hline
   Conv3D & &Conv3DTranspose \\
   BatchNormalisation & & BatchNormalisation \\
   Activation & & Activation \\
   \hline
\hline
 \end{tabular}
 \caption{The architecture of  our U-Net model. The U-Net is created using the \textsc{Keras} library.  The first column gives the layer ID. The 2nd column gives the type of operations.  Convolution blocks and transposed convolution blocks are applied as a loop of 3. The layers constituting the blocks are listed at the bottom of the table.  Concatenations are referred to layer IDs respectively. The third column gives the output dimension of each layer with the format of (N$_{\rm pix}$, N$_{\rm pix}$, N$_{\rm pix}$, N$_{\rm feature}$). }
\label{tab:architect}
\end{table}


\bibliographystyle{mn2e}
\bibliography{journals,lit} 




\bsp	
\label{lastpage}
\end{document}